\def\DIRvalue{Dumitrescu}
\def\IDvalue{DU}
\def\titlevalue{An introduction to supersymmetric \\[1.5pt]
 field theories in curved space}
\def\authorvalue{Thomas T. Dumitrescu}
\def\shortauthorvalue{\authorvalue}
\def\addressvalue{Department of Physics, Harvard University, Cambridge, MA 02138, USA\\
\tt tdumitre@physics.harvard.edu}
\def\abstractvalue{In this review, we give a pedagogical introduction to a systematic framework for constructing and analyzing supersymmetric field theories on curved spacetime manifolds. The framework is based on the use of off-shell supergravity background fields. We present the general principles, which broadly apply to theories with different amounts of supersymmetry in diverse dimensions, as well as specific applications to~$\mathcal{N} = 1$ theories in four dimensions and their three-dimensional cousins with~$\mathcal{N} = 2$ supersymmetry.}
\def\preprintvalue{}

\ifx\ifLONG\undefined
\documentclass[a4paper,11pt]{article}
\newcommand{\chapterauthor}[1]{
\begin{center}
{\bf \normalsize  #1}
\end{center}
}

\newcommand{\chapteraddress}[1]{
\begin{center}
{ \small \it \addressvalue}
\end{center}
}

\newcommand{\chapterabstract}[1]{
\vspace{\baselineskip}
\begin{center}
\textbf{\small Abstract}
\end{center}
#1}

\newcommand{\chapterheader}{

\chapter[\titlevalue{}  (by \shortauthorvalue)]{\titlevalue}
\label{Chapter\IDvalue}
\chapterauthor{\authorvalue}
\chapteraddress{\addressvalue}
\chapterabstract{\abstractvalue}
\tightmtctrue
\minitoc
}

\newcommand{\documentheader}{
\begin{flushright} \small
  \preprintvalue
 \end{flushright}

\begin{center}
{\bf \Large \titlevalue}
\end{center}

\chapterauthor{\authorvalue}
\chapteraddress{\addressvalue}
\chapterabstract{\abstractvalue}

\medskip

This is a contribution to the review volume ``Localization techniques
in quantum field theories'' (eds. V.~Pestun and M.~Zabzine) which
contains 17 Chapters available at \cite{ContributionSummary}

\tableofcontents
}

\newcommand{\ifvolume}[2]{\ifx\ifLONG\undefined#2\else#1\fi}

\newcommand{\documentfinish}{
\ifx\ifLONG\undefined
\bibliographystyle{bibreview} 
\bibliography{\IDvalue,review}  
\end{document}
\else
\addcontentsline{toc}{section}{References}
\providecommand{\href}[2]{#2}\begingroup\raggedright\begin{thebibliography}{10}

\bibitem{ContributionSummary}
V.~Pestun and M.~Zabzine, eds., {\em Localization techniques in quantum field
  theory}, vol.~xx.
\newblock Journal of Physics A, 2016.
\newblock \href{http://arxiv.org/abs/1608.02952}{{\tt 1608.02952}}.
\newblock \url{https://arxiv.org/src/1608.02952/anc/LocQFT.pdf},
  \url{http://pestun.ihes.fr/pages/LocalizationReview/LocQFT.pdf}.

\bibitem{DUSeiberg:1993vc}
N.~Seiberg, ``{Naturalness versus supersymmetric nonrenormalization
  theorems},'' \href{http://dx.doi.org/10.1016/0370-2693(93)91541-T}{{\em Phys.
  Lett.} {\bf B318} (1993)  469--475},
\href{http://arxiv.org/abs/hep-ph/9309335}{{\tt arXiv:hep-ph/9309335
  [hep-ph]}}.

\bibitem{DUIntriligator:1995au}
K.~A. Intriligator and N.~Seiberg, ``{Lectures on supersymmetric gauge theories
  and electric-magnetic duality},''
  \href{http://dx.doi.org/10.1016/0920-5632(95)00626-5}{{\em Nucl. Phys. Proc.
  Suppl.} {\bf 45BC} (1996)  1--28},
\href{http://arxiv.org/abs/hep-th/9509066}{{\tt arXiv:hep-th/9509066
  [hep-th]}}.

\bibitem{DUWitten:1982df}
E.~Witten, ``{Constraints on Supersymmetry Breaking},''
\href{http://dx.doi.org/10.1016/0550-3213(82)90071-2}{{\em Nucl. Phys.} {\bf
  B202} (1982)  253}.

\bibitem{DUWitten:1988ze}
E.~Witten, ``{Topological Quantum Field Theory},''
\href{http://dx.doi.org/10.1007/BF01223371}{{\em Commun. Math. Phys.} {\bf 117}
  (1988)  353}.

\bibitem{ContributionPZ}
V.~Pestun and M.~Zabzine, ``Introduction to localization in quantum field
  theory,'' {\em Journal of Physics A} {\bf xx} (2016)  000,
  \href{http://arxiv.org/abs/1608.02953}{{\tt 1608.02953}}.

\bibitem{DUFestuccia:2011ws}
G.~Festuccia and N.~Seiberg, ``{Rigid Supersymmetric Theories in Curved
  Superspace},'' \href{http://dx.doi.org/10.1007/JHEP06(2011)114}{{\em JHEP}
  {\bf 06} (2011)  114},
\href{http://arxiv.org/abs/1105.0689}{{\tt arXiv:1105.0689 [hep-th]}}.

\bibitem{DUClosset:2013vra}
C.~Closset, T.~T. Dumitrescu, G.~Festuccia, and Z.~Komargodski, ``{The Geometry
  of Supersymmetric Partition Functions},''
  \href{http://dx.doi.org/10.1007/JHEP01(2014)124}{{\em JHEP} {\bf 01} (2014)
  124},
\href{http://arxiv.org/abs/1309.5876}{{\tt arXiv:1309.5876 [hep-th]}}.

\bibitem{DUBaggerQH}
J.~Bagger and J.~Wess, {\em Supersymmetry and Supergravity}.
\newblock Princeton University Press, 1992.

\bibitem{DUAlvarezGaume:1983ig}
L.~Alvarez-Gaume and E.~Witten, ``{Gravitational Anomalies},''
\href{http://dx.doi.org/10.1016/0550-3213(84)90066-X}{{\em Nucl. Phys.} {\bf
  B234} (1984)  269}.

\bibitem{DUWitten:1988hf}
E.~Witten, ``{Quantum Field Theory and the Jones Polynomial},''
\href{http://dx.doi.org/10.1007/BF01217730}{{\em Commun. Math. Phys.} {\bf 121}
  (1989)  351--399}.

\bibitem{DUCasini:2011kv}
H.~Casini, M.~Huerta, and R.~C. Myers, ``{Towards a derivation of holographic
  entanglement entropy},''
  \href{http://dx.doi.org/10.1007/JHEP05(2011)036}{{\em JHEP} {\bf 05} (2011)
  036},
\href{http://arxiv.org/abs/1102.0440}{{\tt arXiv:1102.0440 [hep-th]}}.

\bibitem{ContributionPU}
S.~Pufu, ``The F-Theorem and F-Maximization,'' {\em Journal of Physics A} {\bf
  xx} (2016)  000, \href{http://arxiv.org/abs/1608.02960}{{\tt 1608.02960}}.

\bibitem{DUMoore:1997pc}
G.~W. Moore and E.~Witten, ``{Integration over the u plane in Donaldson
  theory},'' {\em Adv. Theor. Math. Phys.} {\bf 1} (1997)  298--387,
\href{http://arxiv.org/abs/hep-th/9709193}{{\tt arXiv:hep-th/9709193
  [hep-th]}}.

\bibitem{ContributionTA}
Y.~Tachikawa, ``A brief review of the 2d/4d correspondences,'' {\em Journal of
  Physics A} {\bf xx} (2016)  000, \href{http://arxiv.org/abs/1608.02964}{{\tt
  1608.02964}}.

\bibitem{DUWitten:1994ev}
E.~Witten, ``{Supersymmetric Yang-Mills theory on a four manifold},''
  \href{http://dx.doi.org/10.1063/1.530745}{{\em J. Math. Phys.} {\bf 35}
  (1994)  5101--5135},
\href{http://arxiv.org/abs/hep-th/9403195}{{\tt arXiv:hep-th/9403195
  [hep-th]}}.

\bibitem{DUJohansen:1994aw}
A.~Johansen, ``{Twisting of $N=1$ SUSY gauge theories and heterotic topological
  theories},'' \href{http://dx.doi.org/10.1142/S0217751X9500200X}{{\em Int. J.
  Mod. Phys.} {\bf A10} (1995)  4325--4358},
\href{http://arxiv.org/abs/hep-th/9403017}{{\tt arXiv:hep-th/9403017
  [hep-th]}}.

\bibitem{DUKarlhede:1988ax}
A.~Karlhede and M.~Rocek, ``{Topological Quantum Field Theory and $N=2$
  Conformal Supergravity},''
\href{http://dx.doi.org/10.1016/0370-2693(88)91234-8}{{\em Phys. Lett.} {\bf
  B212} (1988)  51}.

\bibitem{DUNekrasov:2002qd}
N.~A. Nekrasov, ``{Seiberg-Witten prepotential from instanton counting},''
  \href{http://dx.doi.org/10.4310/ATMP.2003.v7.n5.a4}{{\em Adv. Theor. Math.
  Phys.} {\bf 7} (2003) no.~5, 831--864},
\href{http://arxiv.org/abs/hep-th/0206161}{{\tt arXiv:hep-th/0206161
  [hep-th]}}.

\bibitem{DUNekrasov:2003rj}
N.~Nekrasov and A.~Okounkov, ``{Seiberg-Witten theory and random partitions},''
  \href{http://dx.doi.org/10.1007/0-8176-4467-9_15}{{\em Prog. Math.} {\bf 244}
  (2006)  525--596},
\href{http://arxiv.org/abs/hep-th/0306238}{{\tt arXiv:hep-th/0306238
  [hep-th]}}.

\bibitem{DUPestun:2007rz}
V.~Pestun, ``{Localization of gauge theory on a four-sphere and supersymmetric
  Wilson loops},'' \href{http://dx.doi.org/10.1007/s00220-012-1485-0}{{\em
  Commun. Math. Phys.} {\bf 313} (2012)  71--129},
\href{http://arxiv.org/abs/0712.2824}{{\tt arXiv:0712.2824 [hep-th]}}.

\bibitem{DUPestun:2014mja}
V.~Pestun, \href{http://dx.doi.org/10.1007/978-3-319-18769-3_6}{``{Localization
  for $\mathcal {N}=2$ Supersymmetric Gauge Theories in Four Dimensions},''} in
  {\em New Dualities of Supersymmetric Gauge Theories}, J.~Teschner, ed.,
  pp.~159--194.
\newblock 2016.
\newblock
\href{http://arxiv.org/abs/1412.7134}{{\tt arXiv:1412.7134 [hep-th]}}.
\newblock

\bibitem{ContributionHO}
K.~Hosomichi, ``$\mathcal{N}=2$ SUSY gauge theories on $S^4$,'' {\em Journal of
  Physics A} {\bf xx} (2016)  000, \href{http://arxiv.org/abs/1608.02962}{{\tt
  1608.02962}}.

\bibitem{DUCordova:2016xhm}
C.~Cordova, T.~T. Dumitrescu, and K.~Intriligator, ``{Deformations of
  Superconformal Theories},''
\href{http://arxiv.org/abs/1602.01217}{{\tt arXiv:1602.01217 [hep-th]}}.

\bibitem{ContributionRR}
L.~Rastelli and S.~Razamat, ``The supersymmetric index in four dimensions,''
  {\em Journal of Physics A} {\bf xx} (2016)  000,
  \href{http://arxiv.org/abs/1608.02965}{{\tt 1608.02965}}.

\bibitem{DUKomargodski:2010rb}
Z.~Komargodski and N.~Seiberg, ``{Comments on Supercurrent Multiplets,
  Supersymmetric Field Theories and Supergravity},''
  \href{http://dx.doi.org/10.1007/JHEP07(2010)017}{{\em JHEP} {\bf 07} (2010)
  017},
\href{http://arxiv.org/abs/1002.2228}{{\tt arXiv:1002.2228 [hep-th]}}.

\bibitem{DUDumitrescu:2011iu}
T.~T. Dumitrescu and N.~Seiberg, ``{Supercurrents and Brane Currents in Diverse
  Dimensions},'' \href{http://dx.doi.org/10.1007/JHEP07(2011)095}{{\em JHEP}
  {\bf 07} (2011)  095},
\href{http://arxiv.org/abs/1106.0031}{{\tt arXiv:1106.0031 [hep-th]}}.

\bibitem{DUGates:1983nr}
S.~J. Gates, M.~T. Grisaru, M.~Rocek, and W.~Siegel, ``{Superspace Or One
  Thousand and One Lessons in Supersymmetry},'' {\em Front. Phys.} {\bf 58}
  (1983)  1--548,
\href{http://arxiv.org/abs/hep-th/0108200}{{\tt arXiv:hep-th/0108200
  [hep-th]}}.

\bibitem{DUFerrara:1974pz}
S.~Ferrara and B.~Zumino, ``{Transformation Properties of the Supercurrent},''
\href{http://dx.doi.org/10.1016/0550-3213(75)90063-2}{{\em Nucl. Phys.} {\bf
  B87} (1975)  207}.

\bibitem{DUKomargodski:2009pc}
Z.~Komargodski and N.~Seiberg, ``{Comments on the Fayet-Iliopoulos Term in
  Field Theory and Supergravity},''
  \href{http://dx.doi.org/10.1088/1126-6708/2009/06/007}{{\em JHEP} {\bf 06}
  (2009)  007},
\href{http://arxiv.org/abs/0904.1159}{{\tt arXiv:0904.1159 [hep-th]}}.

\bibitem{DUDumitrescu:2010ca}
T.~T. Dumitrescu, Z.~Komargodski, and M.~Sudano, ``{Global Symmetries and
  D-Terms in Supersymmetric Field Theories},''
  \href{http://dx.doi.org/10.1007/JHEP11(2010)052}{{\em JHEP} {\bf 11} (2010)
  052},
\href{http://arxiv.org/abs/1007.5352}{{\tt arXiv:1007.5352 [hep-th]}}.

\bibitem{DUKaku:1978nz}
M.~Kaku, P.~K. Townsend, and P.~van Nieuwenhuizen, ``{Properties of Conformal
  Supergravity},''
\href{http://dx.doi.org/10.1103/PhysRevD.17.3179}{{\em Phys. Rev.} {\bf D17}
  (1978)  3179}.

\bibitem{DUStelle:1978ye}
K.~S. Stelle and P.~C. West, ``{Minimal Auxiliary Fields for Supergravity},''
\href{http://dx.doi.org/10.1016/0370-2693(78)90669-X}{{\em Phys. Lett.} {\bf
  B74} (1978)  330--332}.

\bibitem{DUFerrara:1978em}
S.~Ferrara and P.~van Nieuwenhuizen, ``{The Auxiliary Fields of
  Supergravity},''
\href{http://dx.doi.org/10.1016/0370-2693(78)90670-6}{{\em Phys. Lett.} {\bf
  B74} (1978)  333}.

\bibitem{DUSohnius:1981tp}
M.~F. Sohnius and P.~C. West, ``{An Alternative Minimal Off-Shell Version of
  N=1 Supergravity},''
\href{http://dx.doi.org/10.1016/0370-2693(81)90778-4}{{\em Phys. Lett.} {\bf
  B105} (1981)  353--357}.

\bibitem{DUSohnius:1982fw}
M.~Sohnius and P.~C. West, ``{The Tensor Calculus and Matter Coupling of the
  Alternative Minimal Auxiliary Field Formulation of $N=1$ Supergravity},''
\href{http://dx.doi.org/10.1016/0550-3213(82)90337-6}{{\em Nucl. Phys.} {\bf
  B198} (1982)  493--507}.

\bibitem{DUFreedmanZZ}
D.~Z. Freedman and A.~Van~Proeyen, {\em Supergravity}.
\newblock Cambridge University Press, 2012.

\bibitem{DUFerrara:1988qxa}
S.~Ferrara and S.~Sabharwal, ``{Structure of New Minimal Supergravity},''
\href{http://dx.doi.org/10.1016/0003-4916(89)90167-X}{{\em Annals Phys.} {\bf
  189} (1989)  318--351}.

\bibitem{DUClosset:2012vg}
C.~Closset, T.~T. Dumitrescu, G.~Festuccia, Z.~Komargodski, and N.~Seiberg,
  ``{Contact Terms, Unitarity, and F-Maximization in Three-Dimensional
  Superconformal Theories},''
  \href{http://dx.doi.org/10.1007/JHEP10(2012)053}{{\em JHEP} {\bf 10} (2012)
  053},
\href{http://arxiv.org/abs/1205.4142}{{\tt arXiv:1205.4142 [hep-th]}}.

\bibitem{DUClosset:2012vp}
C.~Closset, T.~T. Dumitrescu, G.~Festuccia, Z.~Komargodski, and N.~Seiberg,
  ``{Comments on Chern-Simons Contact Terms in Three Dimensions},''
  \href{http://dx.doi.org/10.1007/JHEP09(2012)091}{{\em JHEP} {\bf 09} (2012)
  091},
\href{http://arxiv.org/abs/1206.5218}{{\tt arXiv:1206.5218 [hep-th]}}.

\bibitem{DUGerchkovitz:2014gta}
E.~Gerchkovitz, J.~Gomis, and Z.~Komargodski, ``{Sphere Partition Functions and
  the Zamolodchikov Metric},''
  \href{http://dx.doi.org/10.1007/JHEP11(2014)001}{{\em JHEP} {\bf 11} (2014)
  001},
\href{http://arxiv.org/abs/1405.7271}{{\tt arXiv:1405.7271 [hep-th]}}.

\bibitem{DUDiPietro:2014bca}
L.~Di~Pietro and Z.~Komargodski, ``{Cardy formulae for SUSY theories in $d =$ 4
  and $d =$ 6},'' \href{http://dx.doi.org/10.1007/JHEP12(2014)031}{{\em JHEP}
  {\bf 12} (2014)  031},
\href{http://arxiv.org/abs/1407.6061}{{\tt arXiv:1407.6061 [hep-th]}}.

\bibitem{DUGomis:2014woa}
J.~Gomis and N.~Ishtiaque, ``{K{\"a}hler potential and ambiguities in 4d $
  \mathcal{N} $ = 2 SCFTs},''
  \href{http://dx.doi.org/10.1007/JHEP04(2015)169}{{\em JHEP} {\bf 04} (2015)
  169},
\href{http://arxiv.org/abs/1409.5325}{{\tt arXiv:1409.5325 [hep-th]}}.

\bibitem{DUAssel:2014tba}
B.~Assel, D.~Cassani, and D.~Martelli, ``{Supersymmetric counterterms from new
  minimal supergravity},''
  \href{http://dx.doi.org/10.1007/JHEP11(2014)135}{{\em JHEP} {\bf 11} (2014)
  135},
\href{http://arxiv.org/abs/1410.6487}{{\tt arXiv:1410.6487 [hep-th]}}.

\bibitem{DUKnodel:2014xea}
G.~Knodel, J.~T. Liu, and L.~A. Pando~Zayas, ``{On N=1 partition functions
  without R-symmetry},'' \href{http://dx.doi.org/10.1007/JHEP03(2015)132}{{\em
  JHEP} {\bf 03} (2015)  132},
\href{http://arxiv.org/abs/1412.4804}{{\tt arXiv:1412.4804 [hep-th]}}.

\bibitem{DUAssel:2015nca}
B.~Assel, D.~Cassani, L.~Di~Pietro, Z.~Komargodski, J.~Lorenzen, and
  D.~Martelli, ``{The Casimir Energy in Curved Space and its Supersymmetric
  Counterpart},'' \href{http://dx.doi.org/10.1007/JHEP07(2015)043}{{\em JHEP}
  {\bf 07} (2015)  043},
\href{http://arxiv.org/abs/1503.05537}{{\tt arXiv:1503.05537 [hep-th]}}.

\bibitem{DUGomis:2015yaa}
J.~Gomis, P.-S. Hsin, Z.~Komargodski, A.~Schwimmer, N.~Seiberg, and S.~Theisen,
  ``{Anomalies, Conformal Manifolds, and Spheres},''
  \href{http://dx.doi.org/10.1007/JHEP03(2016)022}{{\em JHEP} {\bf 03} (2016)
  022},
\href{http://arxiv.org/abs/1509.08511}{{\tt arXiv:1509.08511 [hep-th]}}.

\bibitem{DUAnselmi:1997am}
D.~Anselmi, D.~Z. Freedman, M.~T. Grisaru, and A.~A. Johansen,
  ``{Nonperturbative formulas for central functions of supersymmetric gauge
  theories},'' \href{http://dx.doi.org/10.1016/S0550-3213(98)00278-8}{{\em
  Nucl. Phys.} {\bf B526} (1998)  543--571},
\href{http://arxiv.org/abs/hep-th/9708042}{{\tt arXiv:hep-th/9708042
  [hep-th]}}.

\bibitem{Cassani:2013dba}
D.~Cassani and D.~Martelli, ``{Supersymmetry on curved spaces and
  superconformal anomalies},''
  \href{http://dx.doi.org/10.1007/JHEP10(2013)025}{{\em JHEP} {\bf 10} (2013)
  025},
\href{http://arxiv.org/abs/1307.6567}{{\tt arXiv:1307.6567 [hep-th]}}.

\bibitem{ContributionMO}
D.~Morrison, ``Gromov-Witten invariants and localization,'' {\em Journal of
  Physics A} (2016)  , \href{http://arxiv.org/abs/1608.02956}{{\tt
  1608.02956}}.

\bibitem{DUKlare:2012gn}
C.~Klare, A.~Tomasiello, and A.~Zaffaroni, ``{Supersymmetry on Curved Spaces
  and Holography},'' \href{http://dx.doi.org/10.1007/JHEP08(2012)061}{{\em
  JHEP} {\bf 08} (2012)  061},
\href{http://arxiv.org/abs/1205.1062}{{\tt arXiv:1205.1062 [hep-th]}}.

\bibitem{DUDumitrescu:2012ha}
T.~T. Dumitrescu, G.~Festuccia, and N.~Seiberg, ``{Exploring Curved
  Superspace},'' \href{http://dx.doi.org/10.1007/JHEP08(2012)141}{{\em JHEP}
  {\bf 08} (2012)  141},
\href{http://arxiv.org/abs/1205.1115}{{\tt arXiv:1205.1115 [hep-th]}}.

\bibitem{DUClosset:2014uda}
C.~Closset, T.~T. Dumitrescu, G.~Festuccia, and Z.~Komargodski, ``{From Rigid
  Supersymmetry to Twisted Holomorphic Theories},''
  \href{http://dx.doi.org/10.1103/PhysRevD.90.085006}{{\em Phys. Rev.} {\bf
  D90} (2014) no.~8, 085006},
\href{http://arxiv.org/abs/1407.2598}{{\tt arXiv:1407.2598 [hep-th]}}.

\bibitem{DUKodairabook}
K.~Kodaira, {\em {Complex Manifolds and Deformation of Complex Structures}}.
\newblock Springer, 1986.

\bibitem{DUSeiberg:1994pq}
N.~Seiberg, ``{Electric - magnetic duality in supersymmetric nonAbelian gauge
  theories},'' \href{http://dx.doi.org/10.1016/0550-3213(94)00023-8}{{\em Nucl.
  Phys.} {\bf B435} (1995)  129--146},
\href{http://arxiv.org/abs/hep-th/9411149}{{\tt arXiv:hep-th/9411149
  [hep-th]}}.

\bibitem{DUKodairasone}
K.~Kodaira, ``Complex structures on $S^1 \times S^3$,'' {\em Proceedings of the
  National Academy of Sciences} {\bf 55} (1966) no.~2, 240--243.

\bibitem{DUAssel:2014paa}
B.~Assel, D.~Cassani, and D.~Martelli, ``{Localization on Hopf surfaces},''
  \href{http://dx.doi.org/10.1007/JHEP08(2014)123}{{\em JHEP} {\bf 08} (2014)
  123},
\href{http://arxiv.org/abs/1405.5144}{{\tt arXiv:1405.5144 [hep-th]}}.

\bibitem{DUNishioka:2014zpa}
T.~Nishioka and I.~Yaakov, ``{Generalized indices for $ \mathcal{N} $ = 1
  theories in four-dimensions},''
  \href{http://dx.doi.org/10.1007/JHEP12(2014)150}{{\em JHEP} {\bf 12} (2014)
  150},
\href{http://arxiv.org/abs/1407.8520}{{\tt arXiv:1407.8520 [hep-th]}}.

\bibitem{DURomelsberger:2005eg}
C.~Romelsberger, ``{Counting chiral primaries in N = 1, d=4 superconformal
  field theories},''
  \href{http://dx.doi.org/10.1016/j.nuclphysb.2006.03.037}{{\em Nucl. Phys.}
  {\bf B747} (2006)  329--353},
\href{http://arxiv.org/abs/hep-th/0510060}{{\tt arXiv:hep-th/0510060
  [hep-th]}}.

\bibitem{DUDolan:2008qi}
F.~A. Dolan and H.~Osborn, ``{Applications of the Superconformal Index for
  Protected Operators and q-Hypergeometric Identities to N=1 Dual Theories},''
  \href{http://dx.doi.org/10.1016/j.nuclphysb.2009.01.028}{{\em Nucl. Phys.}
  {\bf B818} (2009)  137--178},
\href{http://arxiv.org/abs/0801.4947}{{\tt arXiv:0801.4947 [hep-th]}}.

\bibitem{DUKinney:2005ej}
J.~Kinney, J.~M. Maldacena, S.~Minwalla, and S.~Raju, ``{An Index for 4
  dimensional super conformal theories},''
  \href{http://dx.doi.org/10.1007/s00220-007-0258-7}{{\em Commun. Math. Phys.}
  {\bf 275} (2007)  209--254},
\href{http://arxiv.org/abs/hep-th/0510251}{{\tt arXiv:hep-th/0510251
  [hep-th]}}.

\bibitem{DUSen:1985ph}
D.~Sen, ``{Supersymmetry in the Space-time $R \times S^3$},''
\href{http://dx.doi.org/10.1016/0550-3213(87)90033-2}{{\em Nucl. Phys.} {\bf
  B284} (1987)  201}.

\bibitem{DUSamtleben:2012gy}
H.~Samtleben and D.~Tsimpis, ``{Rigid supersymmetric theories in 4d Riemannian
  space},'' \href{http://dx.doi.org/10.1007/JHEP05(2012)132}{{\em JHEP} {\bf
  05} (2012)  132},
\href{http://arxiv.org/abs/1203.3420}{{\tt arXiv:1203.3420 [hep-th]}}.

\bibitem{DULiu:2012bi}
J.~T. Liu, L.~A. Pando~Zayas, and D.~Reichmann, ``{Rigid Supersymmetric
  Backgrounds of Minimal Off-Shell Supergravity},''
  \href{http://dx.doi.org/10.1007/JHEP10(2012)034}{{\em JHEP} {\bf 10} (2012)
  034},
\href{http://arxiv.org/abs/1207.2785}{{\tt arXiv:1207.2785 [hep-th]}}.

\bibitem{DUDumitrescu:2012at}
T.~T. Dumitrescu and G.~Festuccia, ``{Exploring Curved Superspace (II)},''
  \href{http://dx.doi.org/10.1007/JHEP01(2013)072}{{\em JHEP} {\bf 01} (2013)
  072},
\href{http://arxiv.org/abs/1209.5408}{{\tt arXiv:1209.5408 [hep-th]}}.

\bibitem{DUKuzenko:2013uya}
S.~M. Kuzenko, U.~Lindstrom, M.~Rocek, I.~Sachs, and
  G.~Tartaglino-Mazzucchelli, ``{Three-dimensional $\mathcal{N} =$ 2
  supergravity theories: From superspace to components},''
  \href{http://dx.doi.org/10.1103/PhysRevD.89.085028}{{\em Phys. Rev.} {\bf
  D89} (2014) no.~8, 085028},
\href{http://arxiv.org/abs/1312.4267}{{\tt arXiv:1312.4267 [hep-th]}}.

\bibitem{DUClosset:2012ru}
C.~Closset, T.~T. Dumitrescu, G.~Festuccia, and Z.~Komargodski,
  ``{Supersymmetric Field Theories on Three-Manifolds},''
  \href{http://dx.doi.org/10.1007/JHEP05(2013)017}{{\em JHEP} {\bf 05} (2013)
  017},
\href{http://arxiv.org/abs/1212.3388}{{\tt arXiv:1212.3388 [hep-th]}}.

\bibitem{DUBG}
M.~Brunella and {\'E}.~Ghys, ``Umbilical foliations and transversely
  holomorphic flows,'' {\em J. Differential Geom} {\bf 41} (1995) no.~1, 1--19.

\bibitem{DUBrunella}
M.~Brunella, ``On transversely holomorphic flows I,'' {\em Inventiones
  mathematicae} {\bf 126} (1996) no.~2, 265--279.

\bibitem{DUGhys}
{\'E}.~Ghys, ``On transversely holomorphic flows II,'' {\em Inventiones
  mathematicae} {\bf 126} (1996) no.~2, 281--286.

\bibitem{DUKapustin:2009kz}
A.~Kapustin, B.~Willett, and I.~Yaakov, ``{Exact Results for Wilson Loops in
  Superconformal Chern-Simons Theories with Matter},''
  \href{http://dx.doi.org/10.1007/JHEP03(2010)089}{{\em JHEP} {\bf 03} (2010)
  089},
\href{http://arxiv.org/abs/0909.4559}{{\tt arXiv:0909.4559 [hep-th]}}.

\bibitem{DUJafferis:2010un}
D.~L. Jafferis, ``{The Exact Superconformal R-Symmetry Extremizes Z},''
  \href{http://dx.doi.org/10.1007/JHEP05(2012)159}{{\em JHEP} {\bf 05} (2012)
  159},
\href{http://arxiv.org/abs/1012.3210}{{\tt arXiv:1012.3210 [hep-th]}}.

\bibitem{DUHama:2010av}
N.~Hama, K.~Hosomichi, and S.~Lee, ``{Notes on SUSY Gauge Theories on
  Three-Sphere},'' \href{http://dx.doi.org/10.1007/JHEP03(2011)127}{{\em JHEP}
  {\bf 03} (2011)  127},
\href{http://arxiv.org/abs/1012.3512}{{\tt arXiv:1012.3512 [hep-th]}}.

\bibitem{ContributionWI}
B.~Willett, ``Localization on three-dimensional manifolds,'' {\em Journal of
  Physics A} {\bf xx} (2016)  000, \href{http://arxiv.org/abs/1608.02958}{{\tt
  1608.02958}}.

\bibitem{DUHama:2011ea}
N.~Hama, K.~Hosomichi, and S.~Lee, ``{SUSY Gauge Theories on Squashed
  Three-Spheres},'' \href{http://dx.doi.org/10.1007/JHEP05(2011)014}{{\em JHEP}
  {\bf 05} (2011)  014},
\href{http://arxiv.org/abs/1102.4716}{{\tt arXiv:1102.4716 [hep-th]}}.

\bibitem{DUImamura:2011uw}
Y.~Imamura, ``{Relation between the 4d superconformal index and the $S^3$
  partition function},'' \href{http://dx.doi.org/10.1007/JHEP09(2011)133}{{\em
  JHEP} {\bf 09} (2011)  133},
\href{http://arxiv.org/abs/1104.4482}{{\tt arXiv:1104.4482 [hep-th]}}.

\bibitem{DUImamura:2011wg}
Y.~Imamura and D.~Yokoyama, ``{N=2 supersymmetric theories on squashed
  three-sphere},'' \href{http://dx.doi.org/10.1103/PhysRevD.85.025015}{{\em
  Phys. Rev.} {\bf D85} (2012)  025015},
\href{http://arxiv.org/abs/1109.4734}{{\tt arXiv:1109.4734 [hep-th]}}.

\bibitem{DUMartelli:2011fu}
D.~Martelli, A.~Passias, and J.~Sparks, ``{The gravity dual of supersymmetric
  gauge theories on a squashed three-sphere},''
  \href{http://dx.doi.org/10.1016/j.nuclphysb.2012.07.019}{{\em Nucl. Phys.}
  {\bf B864} (2012)  840--868},
\href{http://arxiv.org/abs/1110.6400}{{\tt arXiv:1110.6400 [hep-th]}}.

\bibitem{DUNishioka:2013haa}
T.~Nishioka and I.~Yaakov, ``{Supersymmetric Renyi Entropy},''
  \href{http://dx.doi.org/10.1007/JHEP10(2013)155}{{\em JHEP} {\bf 10} (2013)
  155},
\href{http://arxiv.org/abs/1306.2958}{{\tt arXiv:1306.2958 [hep-th]}}.

\bibitem{DUMartelli:2013aqa}
D.~Martelli and A.~Passias, ``{The gravity dual of supersymmetric gauge
  theories on a two-parameter deformed three-sphere},''
  \href{http://dx.doi.org/10.1016/j.nuclphysb.2013.09.012}{{\em Nucl. Phys.}
  {\bf B877} (2013)  51--72},
\href{http://arxiv.org/abs/1306.3893}{{\tt arXiv:1306.3893 [hep-th]}}.

\bibitem{DUAlday:2013lba}
L.~F. Alday, D.~Martelli, P.~Richmond, and J.~Sparks, ``{Localization on
  Three-Manifolds},'' \href{http://dx.doi.org/10.1007/JHEP10(2013)095}{{\em
  JHEP} {\bf 10} (2013)  095},
\href{http://arxiv.org/abs/1307.6848}{{\tt arXiv:1307.6848 [hep-th]}}.

\bibitem{DUNian:2013qwa}
J.~Nian, ``{Localization of Supersymmetric Chern-Simons-Matter Theory on a
  Squashed $S^3$ with $SU(2)\times U(1)$ Isometry},''
  \href{http://dx.doi.org/10.1007/JHEP07(2014)126}{{\em JHEP} {\bf 07} (2014)
  126},
\href{http://arxiv.org/abs/1309.3266}{{\tt arXiv:1309.3266 [hep-th]}}.

\bibitem{DUTanaka:2013dca}
A.~Tanaka, ``{Localization on round sphere revisited},''
  \href{http://dx.doi.org/10.1007/JHEP11(2013)103}{{\em JHEP} {\bf 11} (2013)
  103},
\href{http://arxiv.org/abs/1309.4992}{{\tt arXiv:1309.4992 [hep-th]}}.

\bibitem{Imbimbo:2014pla}
C.~Imbimbo and D.~Rosa, ``{Topological anomalies for Seifert 3-manifolds},''
  \href{http://dx.doi.org/10.1007/JHEP07(2015)068}{{\em JHEP} {\bf 07} (2015)
  068},
\href{http://arxiv.org/abs/1411.6635}{{\tt arXiv:1411.6635 [hep-th]}}.

\bibitem{DUIntriligator:2003jj}
K.~A. Intriligator and B.~Wecht, ``{The Exact superconformal R symmetry
  maximizes a},'' \href{http://dx.doi.org/10.1016/S0550-3213(03)00459-0}{{\em
  Nucl. Phys.} {\bf B667} (2003)  183--200},
\href{http://arxiv.org/abs/hep-th/0304128}{{\tt arXiv:hep-th/0304128
  [hep-th]}}.

\end{thebibliography}\endgroup

\fi
}

\newcommand{\documentfinishBBL}{
\addcontentsline{toc}{section}{References}
\ifx\ifLONG\undefined
\input{\IDvalue.separate.bbl}
\end{document}
\else
\input{\DIRvalue/\IDvalue.volume.bbl}
\fi
}

\ifx\ifLONG\undefined
\def\volcite#1{Contribution \cite{Contribution#1}}
\else 
\def\volcite#1{Chapter \ref{Chapter#1}}
\fi

\usepackage{pdfpages}
\usepackage{fullpage}
\usepackage{mathrsfs} 
\usepackage{amsmath}
\usepackage{amsfonts}
\usepackage{hyperref}
\hypersetup{colorlinks=true}
\hypersetup{linkcolor=blue}
\hypersetup{citecolor=green}
\hypersetup{urlcolor=black}
\numberwithin{equation}{section}\numberwithin{equation}{section}

\begin{document}
\thispagestyle{empty}
\documentheader\else\cite{DUFestuccia:2011ws}
\chapterheader\fi


\section{Introduction}

A standard tool in quantum field theory (QFT) is to probe the theory with non-dynamical sources, or background fields. The consequences of symmetries can then be systematically analyzed by assigning spurious transformation rules to the background fields. In supersymmetric theories, all sources must therefore reside in multiplets of supersymmetry, or superfields. This constrains the extent to which they can affect protected supersymmetric, or BPS, quantities. A typical example is the effective superpotential in four-dimensional theories with~$\mathcal{N}=1$ supersymmetry, which must be a locally holomorphic function of coupling constants that reside in background chiral superfields~\cite{DUSeiberg:1993vc}. This constraint makes it possible to determine the effective superpotential exactly in a large class of theories; see~\cite{DUIntriligator:1995au} for a classic exposition of this powerful approach to analyzing the dynamics of supersymmetric field theories.

Much recent work has involved placing supersymmetric field theories on a manifold~$\mathcal{M}$ with a non-trivial metric or topology, while preserving some (though generally not all) supercharges.\footnote{~The study of supersymmetric field theories on non-trivial manifolds was pioneered by Witten, see for instance~\cite{DUWitten:1982df,DUWitten:1988ze}.} The partition function~$Z_\mathcal{M}$ on~$\mathcal{M}$ (which may be decorated with suitable background fields or operator insertions) is BPS and can sometimes be computed exactly, e.g.~using supersymmetric localization techniques.\footnote{~The basic idea behind supersymmetric localization is reviewed below; see~\volcite{PZ} for a broader and more detailed exposition.} A systematic approach to constructing and analyzing supersymmetric field theories on curved manifolds~$\mathcal{M}$ was presented in~\cite{DUFestuccia:2011ws}. It extends the principle that all background fields should reside in superfields to the metric~$g_{\mu\nu}$ on~$\mathcal{M}$ by embedding it in an off-shell supergravity multiplet.

The purpose of this review is twofold: first, to outline in broad strokes the supergravity-based approach of~\cite{DUFestuccia:2011ws}, which is very general and applies to all supersymmetric field theories. Second, to present some applications to four-dimensional~$\mathcal{N}=1$ theories (section~\ref{DUsec:4d}) and their three-dimensional cousins with~$\mathcal{N}=2$ supersymmetry (section~\ref{DUsec:3d}). These examples illustrate the general framework and showcase its utility for deriving exact results, often without recourse to explicit localization computations, or even a Lagrangian.

\subsection{Background fields and partition functions} 

\label{DUsec:bfpf}

Throughout, background gauge fields coupling to conserved currents will play a crucial role. As an example, consider a theory with a~$U(1)$ flavor symmetry. The corresponding conserved current~$j_\mu$ can be coupled to a background gauge field~$a_\mu$, 
\begin{equation}
\label{DUajlag}
\Delta \mathscr{L} = a^\mu j_\mu + \mathcal{O}(a^2)~.
\end{equation}
The~$\mathcal{O}(a^2)$ seagull terms are tuned to ensure invariance of the Lagrangian under gauge transformations of~$a_\mu$, which enforces current conservation, $\partial^\mu j_\mu = 0$. Small field variations around $a_\mu = 0$ are captured by correlation functions of~$j_\mu$ in the undeformed theory.

Every relativistic QFT possesses a conserved, symmetric stress tensor~$T_{\mu\nu}$. (If the theory is also conformally invariant, then~$T_{\mu\nu}$ can be chosen such that~$T^\mu_\mu =0$.) The appropriate source is a background spacetime metric~$g_{\mu\nu}$. Depending on the signature of spacetime, it may be a Lorentzian or a Riemannian metric. Below, we will mostly discuss field theories on compact, Euclidean spacetime manifolds, which require a Riemannian~$g_{\mu\nu}$. Around flat space, $g_{\mu\nu} = \delta_{\mu\nu}$, the theory couples to a metric deformation~$\Delta g_{\mu\nu}$ via the stress tensor,\footnote{~Unless stated otherwise, we follow the conventions of~\cite{DUClosset:2013vra}. Whenever possible, they coincide with those of~\cite{DUBaggerQH}.}
\begin{equation}
\label{DUflatspace}
g_{\mu\nu} = \delta_{\mu\nu} + \Delta g_{\mu\nu}~, \qquad \Delta \mathscr{L} = - \frac{1}{2} \, \Delta g^{\mu\nu} \, T_{\mu\nu} + \mathcal{O}\left(\Delta g^2\right)~.
\end{equation}
Here the indices are raised and lowered using the flat metric~$\delta_{\mu\nu}$. When the perturbation~$\Delta g_{\mu\nu}$ is small, its effect is captured by correlation functions of~$T_{\mu\nu}$ in flat space. The conservation equation~$\partial^\mu T_{\mu\nu} = 0$ is enforced by choosing the~$\mathcal{O}\left(\Delta g^2\right)$ gravitational seagull terms so that the Lagrangian is invariant under diffeomorphisms that also act on the background metric~$g_{\mu\nu}$. Such a diffeomorphism-invariant Lagrangian can then be studied on an arbitrary Riemannian manifold~$\mathcal{M}$, which may be curved or possess non-trivial topology.\footnote{~Additional care is required if the field theory has gravitational anomalies~(see for instance~\cite{DUAlvarezGaume:1983ig}).}

The stress tensor is not unique: it can be redefined by improvement terms, such as 
\begin{equation} 
\label{DUtimp}
T'_{\mu\nu} = T_{\mu\nu} + \left(\partial_\mu \partial_\nu - \delta_{\mu\nu} \partial^2\right) \mathcal{O}~,
\end{equation}
where~$\mathcal{O}$ is a well-defined scalar operator. Both~$T_{\mu\nu}$ and~$T'_{\mu\nu}$ are acceptable stress tensors: they are symmetric, conserved, and integrate to the momentum operators~$P_\mu$. Consequently, we can use either one to place the theory in curved space. The improvement terms in~\eqref{DUtimp} then give rise to curvature couplings,
\begin{equation} 
\label{DUriccicoup}
\mathscr{L}' = \mathscr{L} - \frac{1}{2} R[g] \mathcal{O}~,
\end{equation} 
where~$R[g]$ is the Ricci scalar of the metric~$g_{\mu\nu}$.\footnote{~In the conventions of~\cite{DUBaggerQH}, a round~$S^d$ of radius~$r$ has constant negative scalar curvature~$R = - \frac{d(d-1)}{r^2}$.} More general improvements can involve a four-index tensor~$\mathcal{O}_{\mu\nu\rho\lambda}$ that couples to the full Riemann tensor~$R_{\mu\nu\rho\lambda}$. We can also modify the Lagrangian by adding local, diffeomorphism-invariant terms that only involve the background metric. These do not change the correlation functions of~$T_{\mu\nu}$ at separated points, but they can give rise to contact terms at coincident points. 

Given a QFT on a manifold~$\mathcal{M}$, it is interesting to study its partition function,
\begin{equation} 
\label{DUpartfndef}
Z_\mathcal{M}\left[\, g_{\mu\nu} \, , \, a_\mu \, , \, \ldots \,\right] = \int \mathcal{D} \Psi \, e^{- \int \mathscr{L}_\mathcal{M} \left[\, \Psi \,; \, g_{\mu\nu} \, , \, a_\mu \, , \, \ldots \,\right]}~.
\end{equation} 
In addition to the metric~$g_{\mu\nu}$ on~$\mathcal{M}$, we can also couple a background gauge field~$a_\mu$ to every flavor current of the theory, as in~\eqref{DUajlag}. The ellipses in~\eqref{DUpartfndef} denote other background fields. Below, we will see that supersymmetric theories are naturally equipped with a variety of other background fields that must be considered in conjunction with~$g_{\mu\nu}$ and~$a_\mu$.  In general, $Z_\mathcal{M}$ suffers from IR and UV divergences. The IR divergences can often be cured by taking~$\mathcal{M}$ to be a compact manifold.\footnote{~This is not sufficient to ensure that~$Z_\mathcal{M}$ is IR finite, since the integral in~\eqref{DUpartfndef} may have bosonic zero modes even if~$\mathcal{M}$ is compact.} As in flat space, the UV divergences are regulated by introducing a short-distance cutoff. The resulting dependence of~$Z_\mathcal{M}$ on the regularization scheme is captured by local counterterms in the background fields. In UV-complete quantum field theories, only finitely many such counterterms are needed. Given a set of background fields, the possible counterterms can be enumerated once and for all. If the regulator preserves certain symmetries, e.g.~diffeomorphisms, the counterterms must also respect these symmetries.

The scheme-independent part of the partition function~$Z_\mathcal{M}$ captures the universal long-distance physics of the QFT. For instance, the functional dependence of~$Z_\mathcal{M}$ on the sources $g_{\mu\nu}, a_\mu,$ etc.~encodes correlation functions of the corresponding local operators~$T_{\mu\nu}, j_\mu$ etc.~on~$\mathcal{M}$. Partition functions can also detect non-local degrees of freedom, which are activated by the topology of~$\mathcal{M}$. A typical example is Chern-Simons theory on a three-manifold, which possesses no local operators but leads to non-trivial partition functions~\cite{DUWitten:1988hf}. 

In conformal field theories (CFTs), the conformal symmetry can be used to relate properties of the theory on different manifolds. A typical example is the operator-state correspondence, which identifies states on~$S^{d-1} \times \mathbb{R}$ in Hamiltonian (i.e.~radial) quantization with local operators. Similarly, correlation functions of local operators on~$\mathbb{R}^d$ are conformally related to correlation functions on~$S^d$, where the IR fluctuations of the CFT are naturally regulated by the finite spacetime volume. Note that conformal symmetry fixes the improvement terms~\eqref{DUtimp}, and hence the curvature couplings~\eqref{DUriccicoup}, by singling out a preferred, traceless stress tensor.

A quantity that has received much recent attention is the entanglement entropy. For the special case of vacuum entanglement across a spherical entangling surface in a CFT, the entanglement entropy can be obtained from the partition function~$Z_{S^d}$ on a round sphere~\cite{DUCasini:2011kv}. More precisely, the statement applies to the universal, scheme-independent parts of both quantities. These can in turn be used to define a quantity that is known (in~$1 \leq d \leq 4$ dimensions) or believed to decrease monotonically under renormalization-group (RG) flow (see for instance~\volcite{PU} and references therein). 

\subsection{Supersymmetric theories}

\label{DUsec:susyth}

As is the case for most observables in interacting QFTs, the partition functions discussed in section~\ref{DUsec:bfpf} are generally not (exactly) computable. The situation is better in supersymmetric theories: BPS observables, which are annihilated by some of the supercharges, are often tightly constrained; in favorable situations, they can even be determined exactly.

Placing supersymmetric field theories on a non-trivial manifold~$\mathcal{M}$ with a curved metric~$g_{\mu\nu}$ generally breaks all flat-space supercharges. Intuitively, this can be understood from the linearized coupling~\eqref{DUflatspace} of the stress tensor to the background metric~$g_{\mu\nu}$, since~$T_{\mu\nu}$ is not a BPS operator, i.e.~$[Q, T_{\mu\nu}] \neq  0$ for every flat-space supercharge~$Q$. More precisely, placing a flat-space theory on~$\mathcal{M}$ by minimally coupling it to the metric~$g_{\mu\nu}$ leads to a curved-space supercharge for each covariantly constant spinor~$\zeta$ on~$\mathcal{M}$,
\begin{equation} 
\label{DUccsonm}
\nabla_\mu \zeta = 0~.
\end{equation} 
This equation is very restrictive. For instance, the only compact four-manifolds that admit covariantly constant spinors are flat tori~$T^4$ and K3 surfaces with Ricci-flat K\"ahler metrics. Similar statements apply to background flavor gauge fields~$a_\mu$, which typically break supersymmetry because the associated flavor current~$j_\mu$ is not a BPS operator. A notable exception occurs for flat connections, which can always be turned on without breaking supersymmetry.\footnote{~This is not true for flat~$R$-symmetry background gauge fields, which can break supersymmetry.}

In this review we will follow~\cite{DUFestuccia:2011ws} and explain how the condition~\eqref{DUccsonm} can be relaxed in a systematic way. Consequently, some supersymmetry can be preserved for a much larger class of manifolds~$\mathcal{M}$ and background fields~$g_{\mu\nu}, a_\mu$. If~$\mathcal{M}$ does not admit covariantly constant spinors, 
this is achieved by coupling the flat-space field theory to background fields in a special, non-minimal way. As we will see, a crucial role is played by additional background fields that are necessarily present in supersymmetric theories. The resulting curved-space Lagrangian~$\mathscr{L}_\mathcal{M}$ is invariant under the action of one or several supercharges, whose algebra may be deformed. The corresponding spinor parameters satisfy equations that generalize~\eqref{DUccsonm}. 

Under favorable conditions, the partition function~$Z_\mathcal{M}$ of a supersymmetric field theory on a curved manifold~$\mathcal{M}$ can be computed exactly using supersymmetric localization. (See~\volcite{PZ} for an overview with references.) The theory is frequently assumed to have a presentation in terms of fields and a Lagrangian. In the simplest case, the curved-space Lagrangian~$\mathscr{L}_\mathcal{M}$ is invariant under a nilpotent supercharge~$Q$, i.e.~$Q^2 = 0$, which can be used to deform the path integral expression~\eqref{DUpartfndef} for the partition function while preserving~$Q$,
\begin{equation}
\label{DUlocpart}
Z_\mathcal{M}(t) = \int \mathcal{D} \Psi \, e^{-\int \mathscr{L}_\mathcal{M} + t \{Q, \mathcal{O}\}}~,
\end{equation} 
for some fermionic operator~$\mathcal{O}$. In order to ensure that the deformed action in the exponent of~\eqref{DUlocpart} is~$Q$-invariant for every value of~$t$, it is convenient (but not necessary) to realize the supercharge~$Q$ off shell. The variation of~$Z_\mathcal{M}(t)$ with respect to the parameter~$t$ vanishes, because the change in the integrand is~$Q$-exact,\footnote{~This argument requires the path integral to converge sufficiently rapidly so that it is legitimate to integrate by parts in field space. See~\cite{DUMoore:1997pc} for a detailed discussion of some examples where this assumption breaks down.}
\begin{equation} 
\label{DUzvar}
\frac{d}{d t} Z_\mathcal{M}(t) = \langle \left\{Q, \mathcal{O}\right\}\rangle = 0~.
\end{equation} 
This shows that~$Z_\mathcal{M} = Z_\mathcal{M}(0)$ can be computed by evaluating~\eqref{DUlocpart} for any choice of~$t$, including~$t \rightarrow \infty$. For suitable choices of the operator~$\mathcal{O}$, this limit localizes the path integral to semiclassical field configurations, with~$t^{-1} \rightarrow 0$ playing the role of Planck's constant. The semiclassical saddle points depend on the choice of~$Q$ and~$\mathcal{O}$, i.e.~they are typically not saddle points of the undeformed theory with Lagrangian~$\mathscr{L}_\mathcal{M}$. 

The bulk of this review volume is dedicated to explicit localization computations of supersymmetric partition functions~$Z_\mathcal{M}$, perhaps in the presence of additional insertions (see~\volcite{PZ} and references therein). The techniques and results reviewed below serve as a basis for such calculations. In particular, we will address the following questions:

\begin{itemize}

\item[1.)] When and how can a supersymmetric field theory  be placed on a curved manifold~$\mathcal{M}$ while preserving some supersymmetry?

\item[2.)] What additional data does the resulting supersymmetric Lagrangian~$\mathscr{L}_\mathcal{M}$ on~$\mathcal{M}$ depend on, beyond the data that was already present in flat space?

\item[3.)] How does supersymmetry constrain the dependence of the partition function~$Z_\mathcal{M}$ on this data? 

\end{itemize}

\noindent As we will see, these questions can be answered within a uniform, largely model-independent framework, which crucially relies on supersymmetry, but not explicit localization computations. In fact, most of the results reviewed below do not require a Lagrangian description of the field theory.\footnote{~See~\volcite{TA} for some examples of localization calculations in non-Lagrangian theories.} Before outlining the general framework in section~\ref{DUsec:overviewgf}, we will examine a few representative examples of supersymmetric field theories in non-trivial backgrounds. 

The only way to preserve all flat-space supercharges on a compact manifold~$\mathcal{M}$ without turning on any background fields other than the metric is to take~$\mathcal{M}$ to be a flat torus~$T^d$, with periodic boundary conditions for fermions. The corresponding partition function~$Z_{T^d}$ is the Witten index~\cite{DUWitten:1982df}, which counts the supersymmetric vacua of the theory on~$T^{d-1} \times \mathbb{R}$, weighted by their fermion number.\footnote{~The Witten index may be ill defined if there are bosonic zero modes that are not lifted when the flat-space theory is compactified on a torus.}

As was already discussed around~\eqref{DUccsonm} above, a covariantly constant spinor leads to a supercharge on~$\mathcal{M}$, but such spinors only exist for very special choices of~$\mathcal{M}$, such as Calabi-Yau manifolds. A more general prescription for preserving supersymmetry, which applies to a larger class of manifolds, is known as twisting~\cite{DUWitten:1988ze}: assume that the supersymmetric theory has a continuous~$R$-symmetry~$G_R$, and that the Riemannian holonomy group of the metric~$g_{\mu\nu}$ on~$\mathcal{M}$ is~$G_\text{hol}$. If a given flat-space supercharge~$Q$ is a singlet under the diagonal subgroup~$(G_R \times G_\text{hol}) \, |_\text{diag}$, then~$Q$ can be preserved on~$\mathcal{M}$. (In flat space, the holonomy group~$G_\text{hol}$ acts via Euclidean rotations.) A prototypical example is topologically twisted~$\mathcal{N}=2$ Yang-Mills theory on an oriented Riemannian four-manifold~\cite{DUWitten:1988ze}. Here~$G_R = SU(2)_R$ and~$G_\text{hol} = SO(4) = SU(2) \times SU(2)$. One of the~$SU(2)$ factors of~$G_\text{hol}$ is twisted by the~$SU(2)_R$ symmetry to yield a single scalar supercharge on~$\mathcal{M}$, which can be used to show that the partition function~$Z_\mathcal{M}$ is independent of the metric~$g_{\mu\nu}$ on~$\mathcal{M}$. For this reason, the twist is referred to as topological. However, not all twists give rise to topological theories. For instance, four-dimensional~$\mathcal{N}=1$ theories with a~$U(1)_R$ symmetry can be twisted on an arbitrary K\"ahler surface~$\mathcal{M}$, for which~$G_\text{hol} = U(2)$~\cite{DUWitten:1994ev,DUJohansen:1994aw}. Now the twisted theory depends on the complex structure of~$\mathcal{M}$, and hence it is not topological. 

Twisted theories are often described by performing a field redefinition to variables that are adapted to the geometric structure that underlies the twist. For instance, topologically twisted~$\mathcal{N}=2$ theories can be described by fields that are differential forms on~$\mathcal{M}$,  while holomorphically twisted~$\mathcal{N}=1$ theories on a K\"ahler surface~$\mathcal{M}$ lead to fields that are complex~$(p,q)$ forms on~$\mathcal{M}$. However, the twisting procedure can also be implemented by coupling the original, untwisted supersymmetric field theory to a background~$R$-symmetry gauge field~$A_\mu^{(R)}$, which is tuned to cancel part of the spin connection~\cite{DUKarlhede:1988ax,DUJohansen:1994aw}. The preserved supercharge on~$\mathcal{M}$ is parametrized by an~$R$-charged spinor~$\zeta$ that satisfies,
\begin{equation} 
\label{DUccsp}
\big(\nabla_\mu - i A_\mu^{(R)} \big) \zeta = 0~,
\end{equation} 
which generalizes~\eqref{DUccsonm}. 

Much recent activity has revolved around supersymmetric field theories on backgrounds that go beyond the basic twisting paradigm. Two prototypical examples of such backgrounds arose in the study of four-dimensional~$\mathcal{N}=2$ theories with an~$SU(2)_R$ symmetry. (We will encounter additional examples below.) The first is the~$\Omega$-background of~\cite{DUNekrasov:2002qd,DUNekrasov:2003rj}, which can be viewed as an equivariant deformation of the topological twist on~$\mathbb{R}^4 = \mathbb{R}^2_{\varepsilon_1} \times \mathbb{R}^2_{\varepsilon_2}$ by an isometry that rotates two orthogonal~$\mathbb{R}^2$ planes inside~$\mathbb{R}^4$. The rotation angles are determined by the equivariant parameters~$\varepsilon_{1,2}$. This background preserves more supercharges than the topological twist, and the corresponding partition function~$Z_\Omega$ explicitly depends on~$\varepsilon_{1,2}$, as well as some flat-space coupling constants, in a complicated and interesting way. The second example is a background on a round~$S^4$, which preserves all eight supercharges~\cite{DUPestun:2007rz}. The supersymmetry algebra is deformed to~$OSp(2|4)$, whose bosonic subalgebra contains the~$SO(2)_R$ Cartan subalgebra of the~$SU(2)_R$ symmetry and the~$Sp(4) = SO(5)$ isometries of~$S^4$. The partition function~$Z_{S^4}$ can depend on some flat-space couplings and the radius of the sphere. See~\cite{DUPestun:2014mja} and~\volcite{HO} for a review of these two backgrounds and some of their applications.

\subsection{Overview of the formalism}

\label{DUsec:overviewgf}

As was noted at the beginning of section~\ref{DUsec:susyth}, the obstruction to preserving supersymmetry on an arbitrary curved manifold~$\mathcal{M}$ is due to the fact that the stress tensor~$T_{\mu\nu}$ is not a  BPS operator. In supersymmetric theories~$T_{\mu\nu}$ resides in a supermultiplet, together with other bosonic and fermionic operators~$\mathcal{J}_B^i$ and~$\mathcal{J}_F^i$. As we will review in section~\ref{DUsec:4dstmult}, the structure of the stress-tensor multiplet reflects very general properties of the field theory (e.g.~the spacetime dimension, the amount of supersymmetry, the presence or absence of possible~$R$-symmetries, or whether the theory is superconformal), but is otherwise largely model independent. Moreover, every supersymmetric field theory must have a stress tensor multiplet, even if the theory is strongly coupled or does not have a Lagrangian description.

The bosonic superpartners~$\mathcal{J}_B^i$ of the stress tensor can be coupled to suitable bosonic background fields~$\mathcal{B}_B^i$ and added to the Lagrangian~\eqref{DUflatspace},
\begin{equation} 
\label{DUgenlagdef}
\Delta \mathscr{L} = - \frac{1}{2} \, \Delta g^{\mu\nu} \, T_{\mu\nu} + \sum_i \mathcal{B}_B^i \mathcal{J}_B^i + \left(\text{seagull terms}\right)~,
\end{equation} 
where we casually refer to all higher-order terms in the background fields as seagull terms. For special choices of~$\Delta g^{\mu\nu}$ and the other bosonic sources~$\mathcal{B}_B^i$, the deformation~$\Delta \mathscr{L}$ can preserve some supersymmetry, due to cancellations between the supersymmetry transformations of~$T_{\mu\nu}$ and~$\mathcal{J}_B^i$. At higher order, we must also ensure supersymmetry of the seagull terms, which can lead to additional conditions.\footnote{~A well-known example arises in four-dimensional~$\mathcal{N}=2$ theories with a continuous flavor symmetry~$G$. We can turn on complex mass parameters~$m$ that are valued in the (complexified) Lie algebra of~$G$. At linear order, all such~$m$ are supersymmetric, but at quadratic order supersymmetry requires that~$\left[m, m^\dagger\right] = 0$. See~\cite{DUCordova:2016xhm} for a recent discussion with references.} 

Following~\cite{DUSeiberg:1993vc}, it was explained in~\cite{DUFestuccia:2011ws}\ that the constraints of supersymmetry on the bosonic sources~$g_{\mu\nu} \,, \,\mathcal{B}_B^i$ are best understood by embedding them into a supermultiplet. Their fermionic superpartners~$\mathcal{B}_F^i$\,, which source the operators~$\mathcal{J}_F^i$ in the stress-tensor multiplet, are set to zero in the Lagrangian~\eqref{DUgenlagdef}. As was emphasized in~\cite{DUFestuccia:2011ws}, the sources must reside in an off-shell supergravity multiplet, because they are non-dynamical background fields that couple to the stress-tensor supermultiplet. This construction can be viewed as a rigid limit of dynamical off-shell supergravity, where the fluctuations of the supergravity fields are frozen by scaling the Planck mass to infinity, $M_p \rightarrow \infty$. We will therefore refer to this construction of supersymmetric field theories on~$\mathcal{M}$ as rigid supersymmetry. 

The requirement that~$\Delta \mathscr{L}$ in~\eqref{DUflatspace} should preserve a supercharge~$Q$ amounts to the statement that the~$Q$-variation of all fermionic sources should vanish,
\begin{equation} 
\label{DUdeltaqferm}
\delta_Q \mathcal{B}_F^i = 0~.
\end{equation} 
The left-hand side of this equation is a non-trivial bosonic expression, which involves the sources~$g_{\mu\nu} \, , \, \mathcal{B}_B^i$ and the spinor~$\zeta$ that parametrizes the supercharge~$Q$. The equations~\eqref{DUdeltaqferm} simultaneously determine the allowed supersymmetric configurations for the bosonic background fields and the corresponding spinor parameter~$\zeta$. 

Even at this level of generality, we can make the following observations:

\begin{itemize}

\item The fermionic sources~$\mathcal{J}_F^i$ always include at least one background gravitino~$\Psi_\mu$, whose supersymmetry variation takes the schematic form~$\delta_Q \, \Psi_\mu = \nabla_\mu \zeta +~\cdots$~. Imposing~\eqref{DUdeltaqferm} then leads to a differential equation for the spinor parameter~$\zeta$ that generalizes~\eqref{DUccsonm} and~\eqref{DUccsp}. We will follow standard practice and refer to such equations as (generalized) Killing spinor equations. A given configuration of background fields admits multiple supercharges if it satisfies~\eqref{DUdeltaqferm} for each supercharge~$Q$, i.e.~if the Killing spinor equations in this background admit multiple independent solutions. 

\item Both the generalized Killing spinor equations and the rigid supersymmetry algebra on~$\mathcal{M}$ follow from the structure of the background off-shell supergravity multiplet. The rigid supersymmetry algebra is realized as a subalgebra of the (infinite-dimensional) algebra of supergravity gauge transformations. As we will review in section~\ref{DUsec:4d}, a given field theory may admit several inequivalent stress-tensor supermultiplets. In this case it can be coupled to different off-shell supergravities,\footnote{~Under certain conditions, distinct off-shell supergravities may be equivalent on shell, but this will not play a role in our discussion.} which generally lead to inequivalent Killing spinor equations, and hence to different supersymmetric backgrounds.

\item A rigid supersymmetric background is characterized by a full set of bosonic supergravity background fields, i.e.~specifying only the metric does not determine the background. In particular, there are distinct backgrounds  that have the same metric but lead to different partition functions. In general, they may arise from different off-shell supergravities, preserve different amounts of supersymmetry, or lead to different supersymmmetry algebras.

\item In Lorentzian signature, unitarity fixes the reality properties of the  fields in the supergravity multiplet so that the Lagrangian~\eqref{DUgenlagdef} is real. In Euclidean signature, we are free to contemplate background fields that do not satisfy the reality conditions needed for unitarity (more precisely, reflection positivity). This greatly enriches the set of Euclidean backgrounds, and some interesting supersymmetric backgrounds can only be obtained in this way. (We will, however, always assume that the background metric~$g_{\mu\nu}$ is a standard Riemannian metric.) Observables (e.g.~partition functions) computed in such non-unitary backgrounds in general do not possess standard reality properties. Nevertheless, they often encode interesting information about the underlying unitarity field theory.

\item If the flat-space field theory has a Lagrangian description in terms of fields, then the Lagrangian and the supersymmetry transformation rules for the fields in curved space follow from the corresponding formulas in the appropriate matter-coupled off-shell supergravity.\footnote{~The formalism only requires the supergravity fields to be off shell. For explicit computations, it is often convenient to also realize some supercharges off shell in the matter sector (see for instance~\volcite{HO}).} These formulas are universal, i.e.~they apply for arbitrary configurations of the supergravity fields. Once a given supersymmetric background has been found, the Lagrangian and the transformation rules in this background can be obtained by specializing the general formulas. 

\end{itemize}

\noindent It is straightforward to extend the preceding discussion to supersymmetric configurations of bosonic background fields residing in other supermultiplets. Supersymmetry requires the variations of all fermionic sources in the multiplet to vanish, as in~\eqref{DUdeltaqferm}. Below we will apply this to background gauge fields that couple to conserved flavor currents. However, the supergravity multiplet enjoys a special status, since it determines the number of supercharges and their algebra. Activating additional background fields that reside in other supermultiplets may preserve these supercharges, or it may break them to a (possibly trivial) subalgebra. 

\subsection{Outline}

In the remainder of this review, we will illustrate the rigid supersymmetry formalism using $\mathcal{N}=1$ theories in four dimensions (section~\ref{DUsec:4d}) and~$\mathcal{N}=2$ theories in three dimensions (section~\ref{DUsec:3d}). We discuss different stress-tensor and supergravity multiplets, and describe some of the corresponding supersymmetric backgrounds. We explain how to construct supersymmetric Lagrangians on these backgrounds and describe the data they depend on, paying particular attention to the data that originates from the coupling to the curved manifold~$\mathcal{M}$. Finally, we explain to what extent this data can affect the partition function~$Z_\mathcal{M}$. We will mostly focus on theories with a~$U(1)_R$ symmetry, but we also mention some results for theories that do not have such a symmetry. 

We consider two examples in detail: $\mathcal{N}=1$ theories on~$S^3 \times S^1$ and~$\mathcal{N}=2$ theories on a round or squashed~$S^3$. The former background can be used to define an index that tracks supersymmetric operators along RG flows (it is closely related to the superconformal index, see~\volcite{RR}). The latter backgrounds play a crucial role in~$F$-maximization and can be used to compute correlation functions of conserved currents (see~\volcite{PU}). 

\section{Four-dimensional~$\mathcal{N}=1$ theories}

\label{DUsec:4d}

\subsection{Stress-tensor multiplets and off-shell supergravities}

\label{DUsec:4dstmult}

As was explained in section~\ref{DUsec:overviewgf}, the procedure of placing a supersymmetric theory on a curved manifold commences with a choice of stress-tensor supermultiplet in flat space. The different possibilities that can arise in four-dimensional~$\mathcal{N}=1$ theories were described in~\cite{DUKomargodski:2010rb,DUDumitrescu:2011iu}. (See also~\cite{DUGates:1983nr} for an early discussion.) Here we will restrict ourselves to the three most common multiplets. We will describe them in superspace (using the conventions of~\cite{DUBaggerQH}) as well as in components. In all cases, the supersymmetry transformations of the component fields implicitly follow from the superspace description. In section~\ref{DUsec:conpartfun4d} we will explicitly write out some of these transformation rules for theories with a~$U(1)_R$ symmetry. 

\begin{itemize}

\item[1.)] The stress-tensor multiplet of an~$\mathcal{N}=1$ superconformal theory (SCFT) is a real superfield~$\mathcal{J}_\mu$ that satisfies
\begin{equation} 
\label{DUscftmult}
\overline D^{\dot \alpha} \mathcal{J}_{\alpha{\dot \alpha}} = 0~, \qquad \mathcal{J}_{\alpha{\dot \alpha}} = \sigma^\mu_{\alpha{\dot \alpha}} \mathcal{J}_\mu~.
\end{equation} 
The component fields in~$\mathcal{J}_\mu$ are given by
\begin{equation} 
\label{DUscftcomp}
\mathcal{J}_\mu = \left(j_\mu^{(R)}\,, S_{\mu\alpha}\,, T_{\mu\nu}\right)~,
\end{equation} 
where~$j_\mu^{(R)}$ is the superconformal~$U(1)_R$ current, $S_{\mu\alpha}$ is the supersymmetry current, and~$T_{\mu\nu}$ is the stress tensor. All three currents are conserved, and the currents~$S_{\mu\alpha}, T_{\mu\nu}$ are traceless, i.e.~$\overline \sigma^{\mu{\dot \alpha}\alpha} S_{\mu\alpha} = T^\mu_\mu = 0$. 

\item[2.)] The majority of four-dimensional~$\mathcal{N}=1$ theories (with or without an~$R$-symmetry) admit a Ferrara-Zumino (FZ) stress-tensor multiplet~\cite{DUFerrara:1974pz}.\footnote{~The only known exceptions are abelian gauge theories with Fayet-Iliopoulos terms, and their analogues in the context of (gauged) sigma models~\cite{DUKomargodski:2009pc,DUKomargodski:2010rb,DUDumitrescu:2010ca,DUDumitrescu:2011iu}.} The FZ-multiplet is given by a real superfield~$\mathcal{J}_\mu^\text{FZ}$, such that
\begin{equation} 
\label{DUfzmultdef}
\overline D^{\dot \alpha} \mathcal{J}_{\alpha{\dot \alpha}}^{\text{FZ}} = D_\alpha X~, \qquad \overline D_{\dot \alpha} X = 0~,
\end{equation} 
where~$\mathcal{J}_{\alpha{\dot \alpha}}^{\text{FZ}} = \sigma^\mu_{\alpha{\dot \alpha}} \mathcal{J}_\mu^\text{FZ}$, as in~\eqref{DUscftmult}. The component fields in the FZ-multiplet are
\begin{equation} 
\label{DUfzcomp}
\mathcal{J}_\mu^\text{FZ} = \left(j_\mu \,, S_{\mu\alpha}\,, x\,, T_{\mu\nu}\right)~.
\end{equation} 
Here~$j_\mu$ is a non-conserved vector operator, $S_{\mu\alpha}$ is the conserved supersymmetry current, $x$ is a complex scalar, and~$T_{\mu\nu}$ is the conserved, symmetric stress tensor. The chiral superfield~$X$ is the trace submultiplet of the FZ-multiplet,\footnote{~Unlike unitary superconformal multiplets, which possess a unique lowest-weight state, multiplets of Poincar\'e supersymmetry may be reducible (i.e.~they may contain non-trivial submultiplets) without being decomposable into smaller multiplets. See~\cite{DUDumitrescu:2011iu} for a discussion in the context of stress-tensor multiplets.} 
\begin{equation} 
\label{DUxfielddef}
X = \left(x \, ,~\sigma^\mu_{\alpha{\dot \alpha}} \overline S_\mu^{\dot \alpha} \, ,~T^\mu_\mu + i \partial^\mu j_\mu \right)~. 
\end{equation} 
When~$X = 0$, the FZ-multiplet reduces to the superconformal multiplet, as can be seen by comparing~\eqref{DUfzmultdef} and~\eqref{DUscftmult}. In this case the vector operator~$j_\mu$ in the FZ-multiplet becomes the conserved superconformal~$U(1)_R$ current, and~$S_{\mu\alpha}, T_{\mu\nu}$ become traceless.

\item[3.)] Non-conformal theories with a~$U(1)_R$ symmetry possess a stress-tensor multiplet~$\mathcal{R}_\mu$, whose bottom component is the conserved~$R$-current~$j_\mu^{(R)}$. In superspace,
\begin{equation}
\label{DUrmult}
\overline D^{\dot \alpha} \mathcal{R}_{\alpha{\dot \alpha}} = \chi_\alpha~, \qquad \overline D_{\dot \alpha} \chi_\alpha = 0~, \qquad D^\alpha \chi_\alpha = \overline D_{\dot \alpha} \overline \chi^{\dot \alpha}~.
\end{equation} 
The component fields residing in the~$\mathcal{R}$-multiplet are given by
\begin{equation}
\label{DUrmultcomp}
\mathcal{R}_\mu = \left(j_\mu^{(R)}\,, S_{\mu\alpha}\,, T_{\mu\nu}\,, C_{\mu\nu}\right)~. 
\end{equation}
Here~$C_{\mu\nu} = C_{[\mu\nu]}$ is a conserved two-form current, which can give rise to a string charge in the supersymmetry algebra~\cite{DUDumitrescu:2011iu}. The superfield~$\chi_\alpha$, which satisfies the same constraints as an abelian field-strength multiplet, is the trace submultiplet of the~$\mathcal{R}$-multiplet. Setting~$\chi_\alpha = 0$ leads to the superconformal multiplet~\eqref{DUscftmult}. 

\end{itemize}

\medskip

\noindent Some theories have more than one stress-tensor multiplet. For instance, a theory with an FZ-multiplet may possess a~$U(1)_R$ symmetry, in which case it also admits an~$\mathcal{R}$-multiplet. In this case the two multiplets are related by a supersymmetric analogue of the improvement transformation~\eqref{DUtimp} for the stress tensor. 

The off-shell supergravity multiplets that couple to the conformal stress-tensor multiplet, the FZ-multiplet and the~$\mathcal{R}$-multiplet are conformal supergravity~\cite{DUKaku:1978nz}, as well as the old~\cite{DUStelle:1978ye,DUFerrara:1978em} and new~\cite{DUSohnius:1981tp,DUSohnius:1982fw} minimal formulations of off-shell supergravity. (See~\cite{DUFreedmanZZ} for a recent discussion of conformal and old minimal supergravity; additional details on new minimal supergravity can be found in~\cite{DUFerrara:1988qxa}.) In principle, we can use any set of off-shell supergravity fields, as long as the flat-space theory admits the corresponding stress-tensor multiplet. In practice, it is often useful to consider non-conformal supergravity, even if the flat-space theory is conformal. The reason is that, quantum mechanically, even CFTs must be defined using a UV cutoff, which breaks conformal symmetry but can often be chosen to preserve supersymmetry. If the theory is conformal, we expect the non-conformal supergravity fields to decouple as the UV cutoff is taken to infinity. However, some remnants of the regulator, and hence of the non-conformal supergravity fields, may survive:

\begin{itemize}

\item The allowed supersymmetric counterterms that parametrized the UV ambiguities (i.e.~the scheme dependence) of the partition function~$Z_\mathcal{M}$ are governed by the non-conformal supergravity theory that couples to the combined SCFT-regulator system. The non-conformal gravity fields can in principle be decoupled by fine-tuning these counterterms, but in practice one is typically left with an ambiguity parametrized by local counterterms that involve the non-conformal supergravity fields.\footnote{~Relevant counterterms are multiplied by positive powers of the UV cutoff~$\Lambda$, so that they are easily identified and adjusted. It is typically more difficult to isolate the effects of marginal counterterms.} This plays an important role in elucidating the properties of supersymmetric partition functions and interpreting the results of explicit localization computations. See for instance~\cite{DUClosset:2012vg,DUClosset:2012vp,DUGerchkovitz:2014gta,DUDiPietro:2014bca,DUGomis:2014woa,DUAssel:2014tba,DUKnodel:2014xea,DUAssel:2015nca,DUGomis:2015yaa} and references therein for a sampling of the recent literature. 

\item The decoupling of the non-conformal supergravity fields can be spoiled by superconformal anomalies, which cannot (even in principle) be removed by fine-tuning the allowed supersymmetric counterterms. Examples are Weyl anomalies in even dimensions, which render~$T^\mu_\mu \neq 0$ in the presence of certain background fields. Such anomalies are, for instance, discussed in~\cite{DUAnselmi:1997am, Cassani:2013dba,DUGomis:2015yaa}, as well as~\volcite{MO}. A different, global superconformal anomaly in three dimensions was described in~\cite{DUClosset:2012vp}.

\end{itemize}
 
\noindent In light of the above, we will only consider the non-conformal old and new minimal supergravity theories.\footnote{~Even though we will not do so here, it is often convenient to formulate non-conformal supergravity theories as coupled systems consisting of a conformal supergravity multiplet and one or several compensating matter multiplets that can be used to Higgs the conformal symmetry.} Moreover, most of our discussion will focus on the new minimal formulation, because field theories with a~$U(1)_R$ symmetry are typically under better theoretical control.

\subsection{Theories with an~$R$-symmetry}

\label{DUsec:4dthwithrsym}

The coupling of theories with a~$U(1)_R$ symmetry to supergravity background fields proceeds via the~$\mathcal{R}$-multiplet~\eqref{DUrmult} and~\eqref{DUrmultcomp}, whose component fields we repeat here for convenience,
\begin{equation}
\label{DUrmultcompii}
\mathcal{R}_\mu = \left(j_\mu^{(R)} \,, S_{\mu\alpha}\,, T_{\mu\nu}\,, C_{\mu\nu}\right)~. 
\end{equation}
The appropriate background fields reside in the new minimal supergravity multiplet~\cite{DUSohnius:1981tp,DUSohnius:1982fw}, 
\begin{equation}
\label{DUnmsugra}
\mathcal{H}_\mu =\left(A_\mu^{(R)} \,, \Psi_{\mu\alpha} \,, g_{\mu\nu} \,, B_{\mu\nu}\right)~.
\end{equation}
In addition to the metric~$g_{\mu\nu}$ and the gravitino~$\Psi_{\mu\alpha}$, this multiplet contains a~$U(1)_R$ gauge field~$A_\mu^{(R)}$, which couples to the conserved~$R$-current~$j_\mu^{(R)}$, and a two-form gauge field~$B_{\mu\nu}$, which couples to the conserved two-form current~$C_{\mu\nu}$. We will often use the Hodge dual of its field strength, which is a covariantly conserved vector field,\footnote{~The factor of~$i$ in~\eqref{DUvdef} is absent in Lorentzian signature, where both~$B_{\mu\nu}$ and~$V^\mu$ are real.}
\begin{equation}
\label{DUvdef}
V^\mu = \frac{i}{2} \varepsilon^{\mu\nu\rho\lambda} \partial_\nu B_{\rho\lambda}~, \qquad \nabla_\mu V^\mu = 0~.
\end{equation}
The only fermionic field in the new minimal supergravity multiplet~\eqref{DUnmsugra} is the gravitino~$\Psi_{\mu\alpha}$. 

As explained around~\eqref{DUdeltaqferm}, the supersymmetric configurations of the bosonic background fields are determined by setting the supersymmetry variations of the gravitino to zero. In new minimal supergravity, these variations take the following form,
\begin{align}
\label{DUnmgravitinovari}
& \delta \Psi_{\mu\alpha} =  -2 \big(\nabla_\mu - i A_\mu^{(R)}\big) \zeta_\alpha -i  V^\nu \sigma_{\mu\alpha{\dot \alpha}} \overline \sigma_\nu^{{\dot \alpha} \beta} \zeta_\beta~,\\
\label{DUnmgravitinovarii}
& \delta \overline \Psi_{\mu}^{\, \dot \alpha} = - 2 \left(\nabla_\mu + i A_\mu^{(R)}\right) \overline \zeta^{\dot \alpha} + i V^\nu \overline \sigma_\mu^{{\dot \alpha} \alpha} \sigma_{\nu \alpha{\dot \beta}} \overline \zeta^{\dot \beta}~.
\end{align}
These formulas are valid in Lorentzian signature, where the left-handed spinor~$\zeta_\alpha$ of~$R$-charge~$+1$ and the right-handed spinor~$\overline \zeta_{\dot \alpha}$ of~$R$-charge~$+1$ are related by complex conjugation, while~$A_\mu^{(R)}$ and~$V_\mu$ are real. 

In Euclidean signature, the left-handed and right-handed spinors are independent and no longer related by complex conjugation. We will emphasize this by writing tildes instead of bars, e.g.~$\widetilde \zeta_{\dot \alpha}$ instead of~$\overline \zeta_{\dot \alpha}$ and~$\widetilde \sigma_\mu$ instead of~$\overline \sigma_\mu$. (In Euclidean signature, we follow the conventions of~\cite{DUClosset:2013vra}.) Moreover, the Lorentzian reality conditions on~$A_\mu^{(R)}$ and~$V_\mu$ may be relaxed at the expense of unitarity. In general, a supercharge~$Q$ is characterized by a pair~$(\zeta, \widetilde \zeta)$ of left- and right-handed Killing spinors, but in new minimal supergravity we can always consider supercharges~$(\zeta, 0)$ or~$(0, \widetilde \zeta)$ of definite~$R$-charge. (In section~\ref{DUsec:norsym} we will discuss theories without an~$R$-symmetry, where this decomposition of~$(\zeta, \widetilde \zeta)$ is generally not possible.) A supercharge~$Q$ of~$R$-charge~$-1$ corresponds to a Killing spinor~$\zeta$ for which the right-hand side of~\eqref{DUnmgravitinovari} vanishes,
\begin{equation}
\label{DUnmkse}
\big(\nabla_\mu - i A_\mu^{(R)}\big) \zeta = -\frac{i}{2} V^\nu \sigma_\mu \widetilde \sigma_\nu \zeta~.
\end{equation}
Similarly, a supercharge~$\widetilde Q$ of~$R$-charge~$+1$ corresponds to a Killing spinor~$\widetilde \zeta$ for which the right-hand side of~\eqref{DUnmgravitinovarii} vanishes,
\begin{equation}
\label{DUnmksebar}
\big(\nabla_\mu + i A_\mu^{(R)}\big) \widetilde \zeta =  \frac{i}{2} V^\nu \widetilde \sigma_\mu \sigma_\nu \widetilde \zeta~.
\end{equation}
Note that these equations reduce to~\eqref{DUccsp}, which describes twisting, when the background field~$V^\mu$ vanishes. 

As explained in section~\ref{DUsec:overviewgf}, the rigid supersymmetry algebra satisfied by the supercharges~$Q$ or~$\widetilde Q$ descends from the algebra of local supergravity transformations. In new minimal supergravity, this algebra includes local supersymmetry transformations (parametrized by arbitrary spinors~$\zeta, \widetilde \zeta$), as well as diffeomorphisms, local Lorentz transformations, and~$R$-symmetry gauge transformations~\cite{DUSohnius:1981tp,DUSohnius:1982fw}. If we restrict to Killing spinors that satisfy~\eqref{DUnmkse} and~\eqref{DUnmksebar}, this algebra simplifies and reduces to the rigid supersymmetry algebra satisfied by the supercharges~$Q, \widetilde Q$. On a field~$\Phi$ with~$U(1)_R$ charge~$r$ and arbitrary spin, the algebra is given by
\begin{equation}
\label{DUnmalg}
\begin{aligned}
& \{\delta_Q, \delta_{\widetilde Q}\} \Phi = 2 i \mathcal{L}_K' \Phi~, \qquad K^\mu = \zeta \sigma^\mu \widetilde \zeta~,\cr
& \delta_Q^2 \Phi = \delta_{\widetilde Q}^2 \Phi = 0~.
\end{aligned}
\end{equation}
The infinitesimal variations anticommute because we take the spinors~$\zeta, \widetilde \zeta$ to be commuting. It follows from the Killing spinor equations~\eqref{DUnmkse} and~\eqref{DUnmksebar} that~$K^\mu$ is a Killing vector. The operator~$\mathcal{L}_K'$ denotes a modified Lie derivative along~$K$, which is twisted by the~$R$-symmetry,
\begin{equation}
\label{DUlprimedef}
\mathcal{L}_K' \Phi = \mathcal{L}_K \Phi - i r K^\mu \left(A_\mu^{(R)} + \frac{3}{2} V_\mu\right) \Phi~.
\end{equation}
Here~$\mathcal{L}_K$ is the ordinary Lie derivative.\footnote{~Its action on spinors~$\chi_\alpha, \widetilde \chi_{\dot \alpha}$ is given by
$$
\mathcal{L}_K \chi = \nabla_\mu \chi - \frac{1}{2} \nabla_\mu K_\nu \sigma^{\mu\nu} \chi~, \qquad \mathcal{L}_K \widetilde \chi = \nabla_\mu \widetilde \chi - \frac{1}{2} \nabla_\mu K_\nu \widetilde \sigma^{\mu\nu} \widetilde \chi~.
$$
}
Due to the twist, the~$R$-charge can appear on the right-hand side of the supersymmetry algebra, unlike in standard flat-space supersymmetry. 

The solutions to the generalized Killing spinor equations~\eqref{DUnmkse} and~\eqref{DUnmksebar} were analyzed in~\cite{DUFestuccia:2011ws,DUKlare:2012gn,DUDumitrescu:2012ha}, and the conditions for the existence of one or several supercharges were deduced. In particular, it was found that a single supercharge~$Q$ of~$R$-charge~$-1$ exists if and only if~$\mathcal{M}$ is a complex manifold, i.e.~it admits an integrable complex structure~${J^\mu}_\nu$, and~$g_{\mu\nu}$ is a compatible Hermitian metric. Since there is only one supercharge, it follows from~\eqref{DUnmalg} that it must square to zero, i.e.~$\delta_Q^2 = 0$. In section~\ref{DUsec:s3s1ex} we will discuss complex manifolds with topology~$S^3 \times S^1$ that preserve up to four supercharges. 

The Killing spinor~$\zeta$ corresponding to a single supercharge~$Q$ on a complex manifold~$\mathcal{M}$ is simply related to the complex structure~${J^\mu}_{\nu}$ on~$\mathcal{M}$,
\begin{equation}
\label{DUjzetarel}
{J^\mu}_\nu = -\frac{2i}{|\zeta|^2} \zeta^\dagger {\sigma^\mu}_\nu \zeta~.
\end{equation}
The background fields~$A_\mu^{(R)}$ and~$V_\mu$ are essentially determined by~${J^\mu}_\nu$ and the Hermitian metric~$g_{\mu\nu}$. Here we will only quote the formula for~$V^\mu$,
\begin{equation}
\label{DUvviaj}
V^\mu = \frac{1}{2} \nabla_\nu {J^\nu}_\mu~,
\end{equation}
up to a freely adjustable piece that will play no role in our discussion. (See~\cite{DUClosset:2013vra} for additional details, including the formula for~$A_\mu^{(R)}$.) Note that~$V^\mu$ vanishes when~$\mathcal{M}$ is K\"ahler, so that~${J^\mu}_\nu$ is covariantly constant. As discussed around~\eqref{DUccsp}, this is precisely the case that allows for twisting by the~$U(1)_R$ symmetry. Therefore, the supergravity construction reduces to twisting in the appropriate limit, but it is more general. For instance, it allows complex manifolds~$\mathcal{M}$ that are not K\"ahler, such as the~$S^3 \times S^1$ backgrounds discussed in section~\ref{DUsec:s3s1ex}. This is only possible because of the additional field~$V^\mu$ supplied by new minimal supergravity. 

An important fact that carries over from twisting is that the supercharge~$Q$ on the complex manifold~$\mathcal{M}$ transforms as a scalar under holomorphic coordinate changes~\cite{DUDumitrescu:2012ha}. This will play a crucial role in section~\ref{DUsec:conpartfun4d}, where we analyze the dependence of the partition function~$Z_\mathcal{M}$ on the geometry of~$\mathcal{M}$. 

It is straightforward to extend the preceding discussion to background gauge fields~$a_\mu$, which couple to conserved flavor currents~$j_\mu$~\cite{DUClosset:2013vra,DUClosset:2014uda}. Here we will focus on a single~$U(1)$ current. In flat space, it resides in a real linear superfield~$\mathcal{J}$, which satisfies
\begin{equation}
\label{DUflcurr}
D^2 \mathcal{J} = \overline D^2 \mathcal{J} = 0~.
\end{equation}
In components,
\begin{equation}
\label{DUflcurrcomp}
\mathcal{J} = \left(J \, , j_\alpha \, , \overline j_{\dot \alpha} \, , j_\mu\right)~, \qquad \partial^\mu j_\mu = 0~.
\end{equation}
The corresponding background gauge field~$a_\mu$ resides in a vector multiplet~$\mathcal{V}$. In Wess-Zumino gauge,
\begin{equation}
\label{DUvmult}
\mathcal{V} = \left(D \, , \lambda_\alpha \, , \overline \lambda_{\dot \alpha} \, , a_\mu\right)~.
\end{equation}
Here~$D$ is a real auxiliary field and~$\lambda_\alpha$ is the gaugino. In order to determine the allowed supersymmetric configurations of the bosonic background fields~$a_\mu, D$ on a complex manifold~$\mathcal{M}$ with supercharge~$Q$, we follow the same logic as above and set
\begin{equation}
\label{DUdeltaliszero}
\delta_Q \lambda  = i \zeta D + \sigma^{\mu\nu} \zeta f_{\mu\nu} = 0~, \qquad f_{\mu\nu} = \partial_\mu a_\nu - \partial_\nu a_\mu~. 
\end{equation}
This leads to the following constraints,
\begin{equation}
\label{DUsusygf}
f^{0,2} = 0~, \qquad D = - \frac{1}{2} J^{\mu\nu} f_{\mu\nu}~,
\end{equation}
where~$f^{0,2}$ is the anti-holomorphic~$(0,2)$ component of the two-form~$f_{\mu\nu}$. Therefore, supersymmetric background gauge fields are in one-to-one correspondence with holomorphic line bundles over the complex manifold~$\mathcal{M}$.

\subsection{Lagrangians}

As was emphasized in~section~\ref{DUsec:overviewgf}, the rigid supersymmetry approach cleanly separates between the allowed supersymmetric backgrounds and their supersymmetry algebras (which were discussed in section~\ref{DUsec:4dthwithrsym}), and supersymmetric Lagrangians on these backgrounds. These Lagrangians only depend on a choice of background supergravity multiplet, but not on the specific field configuration of the supergravity fields. They can be straightforwardly obtained from the corresponding formulas in new-minimal supergravity~\cite{DUSohnius:1981tp,DUSohnius:1982fw,DUFerrara:1988qxa}. 

Consider, for instance, a free chiral multiplet~$\Phi = (\phi, \psi_\alpha, F)$ of~$R$-charge~$r$, and its conjugate anti-chiral multiplet~$\widetilde \Phi = (\widetilde \phi, \widetilde \psi_{\dot \alpha}, \widetilde F)$ of~$R$-charge~$-r$, with flat-space Lagrangian
\begin{equation}
\label{DUflatchi}
\mathscr{L}_{\mathbb{R}^4} = \partial^\mu \widetilde \phi \partial_\mu \phi - i \widetilde \psi \widetilde \sigma^\mu \partial_\mu \psi - \widetilde F F~.
\end{equation}
The corresponding curved-space Lagrangian in the presence of supergravity background fields is given by~\cite{DUFestuccia:2011ws},
\begin{equation}
\label{DUnlchinmlag}
\mathscr{L}_\mathcal{M} = \mathscr{L}_{\mathbb{R}^4} \big|_{\text{covariant}} + V^\mu \left(i \widetilde \phi \, {\overleftrightarrow D_\mu} \phi + \widetilde \psi \widetilde \sigma_\mu \psi\right) - r \left(\frac{1}{4} R - 3V^\mu V_\mu\right) \widetilde \phi \phi~.
\end{equation}
Here~$D_\mu = \partial_\mu - i r A_\mu^{(R)}$ is the~$R$-covariant derivative, and~$\mathscr{L}_{\mathbb{R}^4} \big|_{\text{covariant}}$ is the covariantization of~\eqref{DUflatchi} with respect to diffeomorphisms and~$R$-symmetry gauge transformations. It describes the minimal coupling of~$\mathscr{L}_{\mathbb{R}^4}$ to background fields. However, supersymmetry requires the presence of additional, non-minimal terms in the Lagrangian~\eqref{DUnlchinmlag}. Moreover, these terms explicitly depend on the~$R$-charge~$r$ of~$\Phi$, i.e.~on the choice of~$\mathcal{R}$-multiplet that was used to couple the flat-space theory to background supergravity. This agrees with the general discussion in section~\ref{DUsec:overviewgf}: the coupling to~$\mathcal{M}$ proceeds through the stress-tensor multiplet and different multiplets lead to different theories in curved space. Here the ability to freely assign any~$R$-charge~$r$ to~$\Phi$ reflects the freedom to choose an~$\mathcal{R}$-multiplet from a continuous family of such multiplets. In other situations the~$R$-charge may be fixed, e.g.~in the presence of a superpotential~$W = \Phi^n$ we must set~$r = \frac{2}{n}$.\footnote{~Note that the curvature coupling~$\sim r R \widetilde \phi \phi$ in~\eqref{DUnlchinmlag} may lead to a tachyonic instability if the curvature~$R$ has a definite sign and~$|r|$ is too large. This is born out in explicit examples, e.g.~some supersymmetric partition functions are only meaningful if the~$R$-charges are restricted to a certain range.} 

The non-minimal terms in~\eqref{DUnlchinmlag} also require a corresponding modification of the supersymmetry transformations,
\begin{equation}
\label{DUdeltachinm}
\begin{aligned}
& \delta \phi = \sqrt 2 \zeta \psi~, \cr
& \delta \psi = \sqrt 2 \zeta F + i \sqrt 2 \sigma^\mu \widetilde \zeta \, \big(\partial_\mu - i r A_\mu^{(R)}\big) \phi~,\cr
& \delta F = \sqrt 2 \widetilde \zeta \widetilde \sigma^\mu \left(\nabla_\mu - i (r-1)A_\mu^{(R)} - \frac{i}{2} V_\mu\right)\psi~,
\end{aligned}
\end{equation}
and similarly for the conjugate fields in the anti-chiral multiplet~$\widetilde \Phi$. Given a solution~$\zeta$ of the Killing spinor equation~\eqref{DUnmkse}, we can substitute the corresponding background fields into the Lagrangian~\eqref{DUnlchinmlag} and verify that it is supersymmetric under~\eqref{DUdeltachinm}, provided we use~\eqref{DUnmkse}. 

Broadly speaking, the curved-space Lagrangian~$\mathscr{L}_\mathcal{M}$ depends on three kinds of data:

\begin{itemize}

\item[1.)] Data that was already present in the flat-space Lagrangian~$\mathscr{L}_{\mathbb{R}^4}$. 

\item[2.)] The choice of~$\mathcal{R}$-multiplet that is used to couple the flat-space theory to supergravity background fields. For a theory with a Lagrangian description, this amounts to a set of~$R$-charge assignments for the fields. 

\item[3.)] Various geometric structures on~$\mathcal{M}$, i.e.~the complex structure~${J^\mu}_\nu$, the Hermitian metric~$g_{\mu\nu}$, and possibly background flavor gauge fields described by holomorphic line bundles over~$\mathcal{M}$. These structures emerge from the Killing spinor equations~\eqref{DUnmkse} and~\eqref{DUnmksebar}, as well as~\eqref{DUdeltaliszero} for background gauge fields. 

\end{itemize}

\noindent We will now explain how supersymmetry constrains the dependence of the partition function~$Z_\mathcal{M}$ on this data, focusing on the curved-space data summarized in~$2.)$ and~$3.)$ above.

\subsection{Constraining the partition function}

\label{DUsec:conpartfun4d}

We can use supersymmetry to constrain the dependence of the partition function~$Z_\mathcal{M}$ on continuous data. The basic idea is to vary the data by a small amount, schematically denoted by~$\Delta \mathcal{M}$,  and check whether the corresponding small change~$\Delta \mathscr{L}_\mathcal{M}$ in the Lagrangian is~$Q$-exact. If this is the case, the partition function does not depend on the deformation, 
\begin{equation}
\label{DUqex}
\Delta \mathscr{L}_{\mathcal{M}} = \left(\Delta \mathcal{M}\right) \{Q, \mathcal{O}\}~, \qquad \Delta Z_\mathcal{M} \sim \langle \{Q, \mathcal{O}\}\rangle = 0~.
\end{equation}
The same logic underlies the localization argument, which was sketched around~\eqref{DUlocpart} and~\eqref{DUzvar}. 

A head-on analysis of this problem is possible~\cite{DUClosset:2014uda}, but it is complicated by the fact that the curved-space Lagrangian and the supersymmetry transformations depend on the continuous data that we would like to vary.  Here we will explain a simple but powerful method for sidestepping these complications, which has the added advantage of not requiring a Lagrangian. The simplification proceeds in two steps:

\begin{itemize}
\item[1.)] If we work around flat space, with a nearly flat metric, then the deformation Lagrangian $\Delta \mathscr{L}_\mathcal{M}$ consists of operators in the stress-tensor multiplet of the flat-space theory, i.e.~the~$\mathcal{R}$-multiplet~\eqref{DUrmultcomp}. The known supersymmetry transformations of these operators can be used to determine which terms in~$\Delta \mathscr{L}_\mathcal{M}$ are~$Q$-exact. 

\item[2.)] These results can be extended to arbitrary complex manifolds by using the fact that the supercharge~$Q$ is a scalar under holomorphic coordinate transformations. 
\end{itemize}

\noindent This logic is standard in the context of topological twisting~(see for instance~\cite{DUWitten:1988ze,DUWitten:1994ev}), where~$Q$ is a scalar under all coordinate changes and a suitably defined stress-tensor~$\hat T_{\mu\nu}$ is~$Q$-exact in flat space, $\hat T_{\mu\nu} = \{Q, \Lambda_{\mu\nu}\}$. This is generally sufficient to ensure that the partition function~$Z_\mathcal{M}$ on any four-manifold~$\mathcal{M}$ does not depend on the metric~$g_{\mu\nu}$. 

Following~\cite{DUClosset:2013vra}, we will now apply this argument to constrain the dependence of the partition function~$Z_\mathcal{M}$ on a complex manifold~$\mathcal{M}$ on the complex structure~${J^\mu}_\nu$ and the Hermitian metric~$g_{\mu\nu}$. To this end, we introduce local holomorphic coordinates~$z^i \; (i =1,2)$, in which the non-zero components of the complex structure and the metric are given by
\begin{equation}
\label{DUcsmetcomp}
{J^i}_j = i {\delta^i}_j~, \qquad  {J^{\overline i}}_{\overline j} = - i {\delta^{\overline i}}_{\overline j}~, \qquad g_{i\overline j}~. 
\end{equation}
In these coordinates, infinitesimal variations~$\Delta {J^\mu}_\nu, \Delta g_{\mu\nu}$ of the complex structure and the metric must satisfy the following constraints, 
\begin{equation}
\label{DUdefcon}
\begin{aligned}
& \Delta {J^i}_j = \Delta {J^{\overline i}}_{\overline j} = 0~, \qquad \partial_{\overline j} \Delta {J^i}_{\overline k} - \partial_{\overline k} \Delta {J^i}_{\overline j} = 0~,\cr
& \Delta g_{i\overline j} = \text{anything}~, \qquad \Delta g_{ij } = \frac{i}{2} \left(\Delta J_{ij} + \Delta J_{ji}\right)~.
\end{aligned}
\end{equation}
The first line ensures that~${J^\mu}_\nu + \Delta {J^\mu}_\nu$ is also an integrable complex structure (at first order in the variation), while the second line is the statement that the deformed metric~$g_{\mu\nu} + \Delta g_{\mu\nu}$ should be Hermitian with respect to the deformed complex structure. Complex structure deformations of the form
\begin{equation}
\label{DUtrivialj}
\Delta {J^i}_{\overline j} = 2 i \partial_{\overline j} \varepsilon^i~,
\end{equation}
are induced by an infinitesimal diffeomorphism parametrized by the vector field~$\varepsilon^\mu$. This leads to a cohomology problem for non-trivial complex structure deformations: they correspond to classes in~$H^{0,1}(\mathcal{M}, T^{1,0}\mathcal{M})$. If~$\mathcal{M}$ is compact (as we are assuming here), this is a finite-dimensional vector space, i.e.~there is a finite number of complex structure moduli. See~\cite{DUKodairabook} for an introduction to the deformation theory of complex manifolds. 

We begin with the linearized couplings of the bosonic operators~\eqref{DUrmultcompii} in the~$\mathcal{R}$-multiplet to the bosonic new minimal supergravity fields~\eqref{DUnmsugra} (this is~\eqref{DUgenlagdef}, specialized to new minimal supergravity),\footnote{~Our operator~$C_{\mu\nu}$ was denoted by~$\frac{i}{4} \varepsilon_{\mu\nu\rho\lambda} \mathcal{F}^{\rho\lambda}$ in~\cite{DUClosset:2013vra}.}
\begin{equation}
\label{DUlinnmcouplings}
\Delta \mathscr{L} = - \frac{1}{2} \Delta g^{\mu\nu} T_{\mu\nu}+ A^{(R) \mu} j_\mu^{(R)} + B^{\mu\nu} C_{\mu\nu}~.
\end{equation}
We can now substitute the deformations~\eqref{DUdefcon} into this formula. (This requires the formula for~$B_{\mu\nu}$ in~\eqref{DUvviaj} and the formula for~$A_\mu^{(R)}$ in~\cite{DUClosset:2013vra}.) We find that
\begin{equation}
\label{DUvarylag}
\Delta \mathscr{L} = - \Delta g^{i \overline j} \mathcal{T}_{i \overline j} - i \sum_j {\Delta J^{\overline i}}_j \mathcal{T}_{\overline j \overline i} + i \sum_j {\Delta J^i}_{\overline j} \left(\mathcal{T}_{ij } + i \partial_j j_i^{(R)}\right)~,
\end{equation}
where we have defined the following (complex) linear combination of operators in the~$\mathcal{R}$-multiplet,
\begin{equation}
\label{DUscripttdef}
\mathcal{T}_{\mu\nu} = T_{\mu\nu} +\frac{1}{4} C_{\mu\nu} -\frac{i}{4} \varepsilon_{\mu\nu\rho\lambda} \partial^\rho j^{(R) \lambda} - \frac{i}{2} \partial_\nu j_\mu^{(R)}~.
\end{equation}

We can now ask whether any of these operators are~$Q$-exact, and hence do not affect the partition function when they appear in~\eqref{DUvarylag}. The only fermionic operators in the~$\mathcal{R}$-multiplet are the supersymmetry current~$S_{\mu\alpha}$ and its conjugate~$\widetilde S_{\mu{\dot \alpha}}$, whose~$Q$-variations are given by 
\begin{equation}
\label{DUqofs}
\{Q, S_{\mu\alpha}\} = 0~, \qquad \{Q, \widetilde S_{\mu{\dot \alpha}} \} = 2 i \left(\widetilde \sigma^\nu \zeta\right)_{\dot \alpha} \mathcal{T}_{\mu\nu}~.
\end{equation}
Using the relation~\eqref{DUjzetarel} between the Killing spinor~$\zeta$ and the complex structure~${J^\mu}_\nu$, it can be shown that the second relation in~\eqref{DUqofs} amounts to the statement that all operators of the form~$\mathcal{T}_{\mu \overline i}$, for any index~$\mu$, are~$Q$-exact. Comparing with~\eqref{DUvarylag} shows that:

\begin{itemize}

\item[1.)] The partition function~$Z_\mathcal{M}$ does not depend on the Hermitian metric~$g_{i\overline j}$.

\item[2.)] The partition function~$Z_\mathcal{M}$ depends on~$\Delta {J^i}_{\overline j}$, but not on its complex conjugate~$\Delta {J^{\overline i}}_j$, i.e.~it is a holomorphic function of the complex structure moduli.\footnote{~Note that~$Z_\mathcal{M}$ cannot depend on trivial deformations that vanish in cohomology, since these are induced by background diffeomorphisms.}

\end{itemize}

\noindent These results lead to the following observations: 

\begin{itemize}

\item Since~$Z_\mathcal{M}$ does not depend on the metric, we can rescale~$g_{i \overline j} \rightarrow \lambda^2 g_{i \overline j}$ for some constant~$\lambda$. This uniform scale transformation can be identified with RG flow, and hence~$Z_\mathcal{M}$ can be computed in the UV or in the deep IR of any non-trivial RG flow. An immediate consequence is that~$Z_\mathcal{M}$ must be invariant under IR dualities, such as Seiberg duality~\cite{DUSeiberg:1994pq}. 

\item The arguments above apply to small (infinitesimal) deformations, and hence they only show that~$Z_\mathcal{M}$ is a locally holomorphic function of the complex structure moduli. There are generally interesting singularities at certain loci in moduli space. Even the metric independence of~$Z_\mathcal{M}$ may only hold for sufficiently small deformations (see for instance~\cite{DUMoore:1997pc}). 

\item We can repeat the preceding analysis for flavor current multiplets. The upshot is that~$Z_\mathcal{M}$ only depends on background gauge fields through the corresponding holomorphic line bundles~\cite{DUClosset:2013vra}. In particular, it is a locally holomorphic function of the bundle moduli. If~$\mathcal{M}$ is compact, there are finitely many of them. 

\end{itemize}

So far we have discussed the dependence of~$Z_\mathcal{M}$ on the geometric structures supplied by the background fields. We can use similar methods to analyze its dependence on the choice of~$U(1)_R$ symmetry that is used to couple the flat-space field theory to~$\mathcal{M}$. A detailed discussion can be found in~\cite{DUClosset:2014uda}. Here we only recall that, in flat space, the~$R$-symmetry is not unique whenever there is an abelian flavor symmetry that can mix with it. However, in a non-trivial background the~$R$-charges may be quantized, and hence not continuously variable (see for instance~\cite{DUDumitrescu:2012ha,DUClosset:2013vra}). Only special classes of complex manifolds allow a continuously variable~$R$-symmetry.\footnote{~The precise condition is that the canonical bundle~$\mathcal{K}$ of the complex manifold~$\mathcal{M}$ must be topologically trivial, i.e.~its Chern class must vanish, $c_1(\mathcal{K}) = 0$.}

\subsection{Example: $S^3 \times S^1$}

\label{DUsec:s3s1ex}

We will now briefly summarize an application of the general results discussed above to complex manifolds with topology~$S^3 \times S^1$. (See~\cite{DUClosset:2013vra} for additional details.) It follows from results of Kodaira~\cite{DUKodairasone} that every such complex manifold must be a primary Hopf surface, which comes in two types. We will focus on a primary Hopf surface of the first type, $\mathcal{M}^{p,q}$, which is defined by the following holomorphic quotient, 
\begin{equation}
\label{DUhopfdef}
\mathcal{M}^{p,q} = \left\{\mathbb{C}^2 -(0,0) \right\} / \left\{(w,z) \sim (p w, q z)\right\}~, \qquad 0 < |p| \leq |q| <1~.
\end{equation}
Here~$p,q$ are complex structure moduli of the Hopf surface. The results summarized in section~\ref{DUsec:conpartfun4d} imply that the partition function~$Z_{\mathcal{M}^{p,q}}$ is a locally holomorphic function of~$p,q$. If there are abelian background gauge fields, it must also be locally holomorphic in the corresponding bundle modulus~$u$. (It can be shown that there is only one such modulus on~$\mathcal{M}^{p,q}$.) Partition functions on Hopf surfaces were directly studied in~\cite{DUAssel:2014paa,DUNishioka:2014zpa} using localization techniques. 

It can be shown~\cite{DUClosset:2013vra} that~$Z_\mathcal{M}(p, q, u)$ coincides with the supersymmetric index~$\mathcal{I}(p, q, u)$ for states on~$S^3 \times \mathbb{R}$ defined in~\cite{DURomelsberger:2005eg} (see also~\cite{DUDolan:2008qi, DUKinney:2005ej,DUFestuccia:2011ws}), with general complex fugacities~$p,q,u$.\footnote{~More precisely, the equality between~$Z_\mathcal{M}$ and~$\mathcal{I}$ holds up to a scheme-independent factor, which arises from anomalies and can be interpreted as a supersymmetric Casimir energy~\cite{DUAssel:2014paa,DUAssel:2015nca}.} If the theory is an SCFT, this index coincides with the superconformal index of~\cite{DUKinney:2005ej}, which counts BPS operators, but in general it is distinct. In particular, it is defined away from the conformal point and can be tracked along RG flows. See~\volcite{RR} for a more detailed discussion.

It is worth commenting on the~$S^3 \times \mathbb{R}$ background of new minimal supergravity that is used to define the index~\cite{DUSen:1985ph,DUFestuccia:2011ws}. It preserves four supercharges that anticommute to an~$SU(2|1)$ superalgebra. The bosonic subalgebra~$SU(2) \times U(1)$ contains one of the~$SU(2)$ factors of the~$SU(2)_\ell \times SU(2)_r$ isometry of~$S^3$, and a~$U(1)$ factor that is a linear combination of time translations along~$\mathbb{R}$ and the~$R$-charge. The supergravity background fields are given by
\begin{equation}
\label{DUindsgra}
ds^2 = d\tau^2 + r^2 d\Omega_3~, \qquad V = \pm \frac{i}{r} d\tau~, \qquad A^{(R)} = - \frac{1}{2} V~.
\end{equation}
Here~$r$ is the radius of the round~$S^3$, and the sign of~$V$ depends on whether the~$SU(2) \subset SU(2|1)$ is identified with~$SU(2)_\ell$ or~$SU(2)_r$. The choice of~$A^{(R)}$ is such that the supercharges are time independent. Note that the background fields are consistent with reflection positivity in Euclidean signature, since the~$\tau$-components of~$V$ and~$A^{(R)}$ are purely imaginary, i.e.~they would be real in Lorentzian signature. The non-conformal index~$\mathcal{I}(p, q, u)$ is defined as the Witten index of the theory on~$S^3 \times \mathbb{R}$ in Hamiltonian quantization,
\begin{equation}
\label{DUromberginddef}
\mathcal{I}(p, q, u) = \text{Tr}_{\mathcal{H}_{S^3}} \left((-1)^F p^{J_\ell + J_r - \frac{R}{2}} q^{J_\ell - J_r - \frac{R}{2}} u^{Q_\text{flavor}}\right)~.
\end{equation}
Here~$\mathcal{H}_{S^3}$ is the Hilbert space of states on~$S^3$, $J_\ell$ and~$J_r$ are the Cartan generators of~$SU(2)_\ell$ and~$SU(2)_r$, $R$ is the~$U(1)_R$ charge, and~$Q_\text{flavor}$ is the~$U(1)$ flavor charge associated with the fugacity~$u$. 

\subsection{Theories without an~$R$-symmetry}

\label{DUsec:norsym}

Theories without a~$U(1)_R$ symmetry do not possess an~$\mathcal{R}$-multiplet, and hence they cannot be coupled to the new minimal supergravity background fields. Consequently, the discussion in the preceding subsections does not apply to them. A prominent example of such a theory is pure~$\mathcal{N}=1$ supersymmetric Yang-Mills theory, where the~$U(1)_R$ symmetry is explicitly broken by an anomaly. However, even theories without an~$R$-symmetry typically posses an FZ-multiplet~\eqref{DUfzcomp}, which can be coupled to the old minimal supergravity background fields~\cite{DUStelle:1978ye,DUFerrara:1978em},
\begin{equation}
\label{DUombg}
\mathcal{H}_\mu = \left(b_\mu \,, \Psi_{\mu\alpha} \,, M \,, \widetilde M \,, g_{\mu\nu}\right)~. 
\end{equation}
Here~$b_\mu$ is a well-defined (i.e.~non-gauge) vector field, and~$M, \widetilde M$ are complex scalars. In Lorentzian signature~$\widetilde M = \overline M$, but in Euclidean signature they may be independent.

The Killing spinor equations that follow from setting the supersymmetry variation of the gravitino~$\Psi_{\mu\alpha}$ to zero are given by~\cite{DUFestuccia:2011ws}
\begin{equation}
\label{DUomsgraks}
\nabla_\mu \zeta = \frac{i}{6} M \sigma_\mu \widetilde \zeta +\frac{i}{3} b_\mu \zeta +\frac{i}{3} b^\nu \sigma_{\mu\nu} \zeta~,
\end{equation}
and a similar equation with~$\zeta \leftrightarrow \widetilde \zeta$, $M \leftrightarrow - \widetilde M$, and~$i \leftrightarrow -i$. Note that, unlike in the new minimal case~\eqref{DUnmkse}, the Killing spinor equation mixes the left- and right-handed spinors~$\zeta$ and~$\widetilde \zeta$, which leads to new backgrounds that cannot arise in new minimal supergravity. 

The supersymmetric backgrounds that satisfy~\eqref{DUomsgraks} were classified in~\cite{DUFestuccia:2011ws,DUSamtleben:2012gy,DULiu:2012bi,DUDumitrescu:2012at}. A simple background that highlights the qualitative differences between the old and new minimal cases is a round~$S^4$ of radius~$r$ with\begin{equation}
\label{DUsfourbg}
M = \widetilde M = -\frac{3 i}{r}~, \qquad b_\mu = 0~.
\end{equation}
Since~$S^4$ is not a complex manifold, it cannot arise as a background in new minimal supergravity. Moreover, the non-zero values for~$M, \widetilde M$ necessarily break the~$R$-symmetry of the field theory, even if it was present in flat space. Finally, note that~$M, \widetilde M$ are not complex conjugates, and hence the background does not respect reflection positivity unless these fields decouple. This happens if the flat-space theory is superconformal, in which case it can be mapped to~$S^4$ by a conformal transformation that preserves unitarity. In a non-conformal theory, the violation of unitarity is necessary in order to avoid a no-go theorem that forbids unitary supersymmetric theories in de Sitter space, and hence reflection positive supersymmetric theories on compact spheres. The~$S^4$ background admits a squashing deformation that only preserves the isometry group~$SO(4) \subset SO(5)$. Unfortunately, neither the round nor the squashed~$S^4$ appear to be amendable to localization calculations (see for instance the recent discussion in~\cite{DUKnodel:2014xea}).

\section{Three-dimensional~$\mathcal{N}=2$ theories}

\label{DUsec:3d}

\subsection{Theories with an~$R$-symmetry on curved manifolds}

Here we briefly sketch extensions of the results summarized in section~\ref{DUsec:4d} to three-dimensional theories with~$\mathcal{N}=2$ supersymmetry. We only discuss theories with a~$U(1)_R$ symmetry. Now the~$\mathcal{R}$-multiplet consists of the following operators~\cite{DUDumitrescu:2011iu},
\begin{equation}
\label{DUtdrmult}
\mathcal{R} = \left(j_\mu^{(R)} \,, S_{\mu\alpha} \,, T_{\mu\nu} \,, j_\mu^{(Z)} \,, J\right)~.
\end{equation}
Here~$j_\mu^{(R)}$ is the~$R$-current, $S_{\mu\alpha}$ is the supersymmetry current, $T_{\mu\nu}$ is the stress tensor, $j_\mu^{(Z)}$ is the central charge current, and~$J$ is a scalar operator. All operators other than~$J$ are conserved currents. The corresponding background supergravity fields constitute the analogue of new minimal supergravity in three dimensions~(see for instance~\cite{DUKuzenko:2013uya} and references therein),
\begin{equation}
\label{DUtdsugra}
\mathcal{H} = \left(A_\mu^{(R)} \,, \Psi_{\mu\alpha} \,, g_{\mu\nu} \,, C_\mu \,, H\right)~.
\end{equation}
Now the condition~$\delta_Q \Psi_{\mu\alpha} = 0$ leads to the following generalized Killing spinor equation for the allowed supersymmetric backgrounds~\cite{DUKlare:2012gn,DUClosset:2012ru},
\begin{equation}
\label{DUtdkse}
\left(\nabla_\mu - A_\mu^{(R)}\right) \zeta = - \frac{1}{2} H \gamma_\mu \zeta + \frac{i}{2} V_\mu \zeta - \frac{1}{2} \varepsilon_{\mu\nu\rho} V^\nu \gamma^\rho \zeta~. 
\end{equation}
Here~$V^\mu = - i \varepsilon^{\mu\nu\rho} \partial_\nu C_\rho$ is the dual field strength of~$C_\mu$ in Euclidean signature. A solution~$\zeta$ to these equations exists if and only if the three-manifold~$\mathcal{M}$ admits a geometric structure known as a transversely holomorphic foliation (THF), and the metric is a compatible transversely Hermitian metric (see~\cite{DUClosset:2013vra} for additional details). This structure is comprised of the following ingredients:
\begin{itemize}

\item[1.)] A nowhere vanishing unit vector field~$\xi^\mu$, which provides a local~$2+1$ decomposition of the manifold~$\mathcal{M}$.

\item[2.)] An integrable complex structure~$J$ on the two-dimensional spaces transverse to~$\xi^\mu$, such that~$J$ is invariant along~$\xi^\mu$, i.e.~$\mathcal{L}_\xi J = 0$.

\end{itemize}

\smallskip
\noindent In the compact case, such manifolds have been classified~\cite{DUBG,DUBrunella,DUGhys}. Topologically, they must be Seifert manifolds or~$T^2$ bundles over~$S^1$. Compact hyperbolic three-manifolds are not allowed.

As is already clear from the definition, manifolds that carry a THF are very similar to complex manifolds. For instance, both admit complex~$(p,q)$ differential forms, a~$\overline \partial$-operator, a corresponding Dolbeault cohomology, and holomorphic line bundles. As in four dimensions, these holomorphic line bundles correspond to supersymmetric configurations of background gauge fields for abelian flavor symmetries. Both a THF, and the holomorphic line bundles over it, generally come in infinite families labled by a finite number of holomorphic moduli. As in the discussion around~\eqref{DUtrivialj}, these moduli (which are finite in number if~$\mathcal{M}$ is compact) correspond to certain~$\overline \partial$-cohomology classes. See section~5 of~\cite{DUClosset:2013vra} for an introduction to THFs and their moduli.

\subsection{Constraining the partition function}

\label{DUsec:conspf3d}

In addition to the flat-space couplings and the choice of~$R$-symmetry, the Lagrangian on~$\mathcal{M}$ now depends on a choice of THF, a transversely Hermitian metric, and holomorphic line bundles corresponding to background flavor gauge fields. Repeating the arguments in section~\ref{DUsec:conpartfun4d} in this case, we find that (see \cite{DUClosset:2013vra} for a detailed discussion):

\begin{itemize}

\item The partition function~$Z_\mathcal{M}$ does not depend on the transversely Hermitian metric. 
\item $Z_\mathcal{M}$ is a locally holomorphic function of the THF moduli. 
\item The partition function depends holomorphically on line bundle moduli corresponding to background flavor gauge fields. 

\end{itemize}

\subsection{Example: round and squashed~$S^3$}

\label{DUsec:3drsqs3}

In~$\mathcal{N}=2$ theories with a~$U(1)_R$ symmetry, the partition function on a round~$S^3$ is computable using supersymmetric localization techniques~\cite{DUKapustin:2009kz,DUJafferis:2010un,DUHama:2010av}  (see also~\volcite{WI}). This result has been generalized to a large variety of squashed spheres, see for instance~\cite{DUHama:2011ea,DUImamura:2011uw,DUImamura:2011wg,DUMartelli:2011fu,DUNishioka:2013haa,DUMartelli:2013aqa,DUAlday:2013lba,DUNian:2013qwa,DUTanaka:2013dca}. These squashed spheres often have the feature that their metric contains arbitrary functions, in addition to various continuous parameters. Explicit localization computations of partition functions on these squashed spheres indicate that:

\begin{itemize}

\item The partition function only depends on the background geometry through a single complex parameter~$b$, known as the squashing parameter. We will therefore denote the partition function by~$Z_{S^3_b}$. 
\item Some deformations of the background fields do not affect~$Z_{S^3_b}$ (i.e.~they do not change~$b$), even though the metric changes. 

\end{itemize}

These observations can be understood using the results of~\cite{DUClosset:2013vra} summarized in section~\ref{DUsec:conspf3d} above.\footnote{~Some of these results (for special backgrounds and theories) were subsequently reproduced from a different point of view in~\cite{Imbimbo:2014pla}. We thank the authors for emphasizing their work to us.} It follows from the classification of~\cite{DUBG,DUBrunella,DUGhys} that the moduli space of THFs on three-manifolds diffeomorphic to~$S^3$ (i.e.~squashed spheres) is one complex dimensional.\footnote{~There is another, isolated branch of the moduli space, which consists of a single point, but it will not be important for us here (see \cite{DUClosset:2013vra} for additional details).} Therefore all squashed-sphere partition functions should only depend on one complex modulus, which can be identified with the squashing parameter~$b$. It also shows that more complicated squashings will not lead to new partition functions. 

Similarly, distinct squashed spheres that give rise to the same value of~$b$ correspond to the same choice of THF, but possibly different transversely Hermitian metrics, which do not affect the partition function.

\subsection{$F$-maximization and correlation functions}

The SUSY theories on~$S^3 \times S^1$ and~$S^3$ discussed in sections~\ref{DUsec:s3s1ex} and~\ref{DUsec:3drsqs3} above explicitly depend on a choice of~$U(1)_R$ symmetry, which affects their curvature couplings. In a superconformal theory, there is a distinguished choice of~$U(1)_R$ symmetry, which resides in the superconformal algebra. In four-dimensional~$\mathcal{N}=1$ theories, it can be determined in flat space using anomalies and~$a$-maximization~\cite{DUAnselmi:1997am,DUIntriligator:2003jj}.

The analogous principle for three-dimensional $\mathcal{N}=2$ theories is $F$-maximization \cite{DUJafferis:2010un}. Since this is the subject of~\volcite{PU}, we will only make a few remarks. Consider the partition function~$Z_{S^3}$ on a round~$S^3$, together with a supersymmetric background gauge field for the conserved flavor current~$j_\mu$. This partition function only depends on one holomorphic line bundle modulus~$u$,
\begin{equation}
\label{DUsthreepf}
Z_{S^3} = e^{-F(u)}~, \qquad F(u) = F(m + i t)~.
\end{equation}
Here~$t \in \mathbb{R}$ controls the mixing of the flavor symmetry with the~$R$-symmetry, while~$m$ is a real mass parameter associated with the flavor symmetry. The fact that the~$m$- and~$t$-dependence of~$F$ descends from a single holomorphic function of~$u$ was first observed in~\cite{DUJafferis:2010un}. A general explanation was given in~\cite{DUClosset:2014uda}. 

Derivatives of the free energy~$F$ with respect to~$t$ compute integrated correlation functions of~$j_\mu$ or its superpartners on~$S^3$. In an SCFT, one-point functions should vanish, so that
\begin{equation}
\label{DUfext}
\partial_t \, \text{Re} F \big|_\text{SCFT} = 0~.
\end{equation}
Surprisingly, the first derivative of the imaginary part~$\text{Im} F$ need not vanish, due to a global superconformal anomaly that can arise in three dimensions~\cite{DUClosset:2012vg,DUClosset:2012vp}.

Taking more derivatives with respect to~$t$ leads to higher-point correlation functions of~$j_\mu$, for instance
\begin{equation}
\label{DUtwopointjj}
\partial_t^2 \, \text{Re} F \big|_\text{SCFT} = -\frac{\pi^2}{2} \,\tau~.
\end{equation}
Here~$\tau$ is the coefficient of the current two-point function at separated points in flat space. In a unitary theory~$\tau$ must be positive, 
\begin{equation}
\label{DUtaudef}
\langle j_\mu(x) j_\nu(0)\rangle =  \frac{\tau}{16 \pi^2} \left(\delta^{\mu\nu} \partial^2 - \partial^\mu  \partial^\nu\right) \frac{1}{x^2}~, \qquad \tau > 0~.
\end{equation}
The conditions in~\eqref{DUfext}, \eqref{DUtwopointjj}, and~\eqref{DUtaudef} amount to the statement of~$F$-maximization, which can be used to solve for the superconformal value~$t = t_*$ of the mixing parameter. Once this value has been found, we can use~\eqref{DUtwopointjj} to compute the value of~$\tau$ in the SCFT. Similarly, we can slightly squash the sphere away from the round point~$b = 1$ to extract the positive coefficient~$C_T >0$ that appears in the stress-tensor two-point function at separated points~\cite{DUClosset:2012ru}, 
\begin{equation}
\label{DUsmallsquash}
C_T \sim \partial_b^2 \, \text{Re} F \big|_{b = 1}~.
\end{equation}

\section*{Acknowledgments}

I am grateful to C.~Closset, G.~Festuccia, Z.~Komargodski, and N.~Seiberg for collaboration on some of the work reviewed here. My work is supported by the Fundamental Laws Initiative at Harvard University, as well as DOE grant DE-SC0007870 and NSF grant PHY-1067976.

\documentfinish